\documentstyle[prl,aps,epsf,amsfonts,amssymb,epsfig]{revtex}

\begin{document}
\draft
\twocolumn[\hsize\textwidth\columnwidth\hsize\csname@twocolumnfalse\endcsname
\title{Optical Properties of Layered Superconductors near the Josephson Plasma 
Resonance}
\author{Ch. Helm$^{1,2}$ and L. N. Bulaevskii$^2$}
\address{$^1$ETH H{\"o}nggerberg, Institut f{\"u}r Theoretische Physik, 
Z{\"u}rich, Switzerland \\
$^2$Los Alamos National Laboratory, Los Alamos, NM 87545, USA \\ }
\date{\today, submitted to PRB}
\maketitle
\begin{abstract}
We study the optical properties of crystals with spatial dispersion and show 
that the usual Fresnel approach becomes invalid near frequencies  
where the group velocity of the wave packets inside the crystal vanishes. 
Near these special frequencies the reflectivity 
depends on the atomic structure of the crystal provided 
that disorder and dissipation are very low.  This is demonstrated explicitly 
by a detailed study of layered superconductors with identical or two different 
alternating junctions in the frequency range near the Josephson plasma 
resonance. Accounting for both 
inductive and charge coupling of the intrinsic junctions, we show that 
multiple modes are excited inside the crystal by the incident light, 
determine their relative amplitude by the microscopic calculation of the 
additional boundary conditions and finally obtain the reflectivity.  
 Spatial dispersion also provides a 
novel method  to stop light pulses, which has possible applications for 
quantum information processing and the artificial creation of event horizons 
in a solid. 
\end{abstract}
]


\section{ Introduction}

The problem of optical properties of  crystals 
with spatial dispersion has remained challenging since the original paper of 
Pekar on the optics of  exciton bands \cite{pekar}.  
Despite considerable effort, the complete theoretical description of the 
optical properties of such systems is still missing 
\cite{pekar2,agr,henn,birman1,birman2,chen,forst}.  

The nontrivial optical features of crystals with a dispersive dielectric 
function $\epsilon(\omega,{\bf k})$ are based on the fact that 
incident light with a given frequency excites several eigenmodes with 
different wave vectors ${\bf k}$.
This poses the 
fundamental problem that the Maxwell boundary conditions, i.e. the 
continuity of the electric and magnetic field 
components parallel to the surface, are insufficient to calculate the 
relative amplitudes of these modes and consequently to describe
physical quantities, such as reflectivity or transmissivity.   
Since the early work of Pekar \cite{pekar,pekar2} and Ginzburg \cite{agr}
this difficulty was usually addressed in a purely 
phenomenological approach by introducing so called additional 
boundary conditions (ABC) for the macroscopic polarization. These ABC are 
motivated physically by the microscopic structure of the surface, but 
the choice of ABC is not universal and may be
controversial, see Ref.~\onlinecite{henn} and Comments to this paper. 
Only the complete solution of the microscopic model can 
determine the dependence of the reflectivity on the microstructure 
unambiguously.  

Such a solution was found recently for the first time for the reflectivity 
near the Josephson plasma resonance (JPR) 
in highly anisotropic layered superconductors \cite{ourletter}, 
which is an interlayer charge oscillation due to the tunneling of 
Cooper pairs and quasiparticles in highly anisotropic layered 
superconductors \cite{kosh,lnb,bul}.  Josephson plasma oscillations inside 
layered superconductor may be excited by the light incident to the surface 
of the crystal in the geometries (a) or (b) shown in Fig.~1. The  
JPR in layered superconductors is the simplest example, 
which illuminates the effects of spatial dispersion and the discrete atomic 
structure on optical properties in strongly anisotropic materials.   
Here we will describe the method of the calculations in \cite{ourletter} 
in more detail, generalize our results for the JPR to different 
geometries, discuss the various transmission and reflection 
coefficients in a finite size sample and point out  perspectives to stop 
light with the help of spatial dispersion. We also stress that the discrete 
atomic structure within the unit cell can have similar effects as  spatial 
dispersion. 

\begin{figure}
\begin{center} 
\epsfig{file=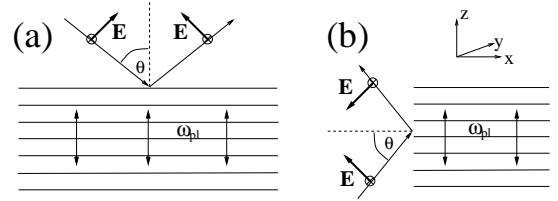,width=0.40\textwidth,clip=,angle=0}
\end{center}
\caption{The geometry of the layered system showing the incident and reflected
light at the surface of incidence
(a) parallel and (b) perpendicular to the layers. Interlayer charge 
oscillations (vertical arrows) are excited by the component of 
the electric field perpendicular to the layers.
\label{geometry}
}
\end{figure}

In the framework of the Lawrence-Doniach model \cite{ld} 
(interlayer Josephson coupling) we can describe both layered superconductors 
with identical intrinsic Josephson junctions (such as Tl-2201 
\cite{tors1,tors2},  
Bi$_2$Sr$_2$CaCu$_2$O$_8$ \cite{bscco}, 
the organic material $\kappa$-(BEDT-TTF)$_2$-Cu(NCS)$_2$  
\cite{organic,dressel}  
or (LaSe)(NbSe$_2)$ \cite{nbse1,nbse2}) and compounds,  where  different 
junctions alternate like in SmLa$_{1-x}$Sr$_{x}$CuO$_{4-\delta}$ 
\cite{marel,marel2,shibata,kakeshita,pim,gorshunov}, Bi-2212/Bi-2201  
\cite{bsccomulti} or atomic scale YBCO/PrBCO superlattices  \cite{ybcomulti}. 
Thereby we take into account not only the dispersion of the plasma mode 
caused by the inductive interaction of currents parallel to the layers, but 
also the $c$-axis dispersion due to charge fluctuations on the layers 
\cite{art,koy,tach,sendai,sandiego,ryndyk1}. 

The JPR is an ideal choice to illustrate the effect of spatial dispersion 
and the atomic structure on optical properties both theoretically and 
experimentally.
First of all, recent optical experiments on the layered superconductor 
SmLa$_{1-x}$Sr$_{x}$CuO$_{4-\delta}$ with T$^*$ crystal structure
showed evidence that the spatial dispersion 
of the Josephson plasmon in the direction perpendicular to the layers 
is important 
\cite{marel,marel2,shibata,kakeshita,pim}. 
For incidence parallel to the layers, see Fig.~1(b) at $\theta=0$,  
two peaks at $\approx 7$ and $\approx 12$ cm$^{-1}$ were observed 
in reflection, which can  be naturally understood as the JPR \cite{marel}
of alternating intrinsic junctions with SmO or LaO in the barriers
between the CuO$_2$-layers 
\cite{marel,marel2,shibata,kakeshita,pim,str}.  The very high ratio of 
the peak intensities, about 20, cannot be explained in  a dispersionless model
\cite{marel3} and it points to a quite strong  $c$-axis
dispersion of the plasma modes due to charge variations
\cite{marel4,ourpaper}. 
Secondly, from the theoretical point of view 
the well established Lawrence-Doniach model \cite{ld}
formulated in terms of finite-difference equations for electromagnetic 
fields and phases of the superconducting order parameter is sufficient to 
provide a complete microscopic description and can be solved analytically.   
Lastly, it is fortunate that the damping due to dissipation is low, because at 
low temperatures the JPR frequency is well below the superconducting gap
and the quasiparticles responsible for dissipation 
are frozen out. Otherwise it would strongly overshadow the effects of
dispersion or the atomic structure as described below. 

Extracting the strength of the $c$-axis dispersion in high 
temperature superconductors is important in its own, as the dynamics of 
Josephson oscillations in layered superconductors is strongly influenced by it
\cite{koy,sendai,sandiego}. It is also intimately 
connected with the electronic compressibility  of the
 superconducting CuO$_2$-layers, which is hard to
measure in situ otherwise and contains unique information about the 
electronic many-body interactions in the layers.

From a more fundamental point of view, we show that in the presence of 
spatial dispersion the conventional Fresnel formulas for reflectivity and 
transmission have to be modified 
substantially near certain frequencies, if both the dissipation and the 
crystal disorder are weak. 
Usually it is assumed that the optical properties of crystals are completely 
determined by average, bulk properties described by a frequency dependent 
 dielectric function $\epsilon (\omega)$, but not by the {\em explicit} 
spatial dispersion 
 (${\bf k}$-dependence) or the specific atomic structure of the crystal
({\em implicit} spatial dispersion).  This is based on the
notion that the wave length of light is much larger than the atomic length
scales  and  therefore light is expected to be influenced 
only by averaged properties of the crystal. Here we will stress out that 
this approach breaks down, if the group velocity,  
${\bf v}_g=\partial \omega({\bf k})/\partial{\bf k}$, of the 
wave packet of the optical excitation with dispersion $\omega({\bf k})$ 
becomes small. 
The physical reason for this breakdown of the macroscopic 
theory is  the appearance of a small effective wave length, 
$\lambda_g=v_g/\omega$, related to the slow motion of the wave packet, 
which can be comparable with the interatomic distance. 

The conditions, when the group velocity becomes small, can be most easily seen
for an isotropic medium described by the dielectric 
function $\epsilon(\omega,k)$.  Then the 
dispersion relation of an optically excited eigenmode is 
$c^2 k^2 = \omega^2 \epsilon ( \omega, k ) = \omega^2 n^2 (\omega, k)$. 
For a transversal wave the implicit derivative of this equation with 
respect to 
$k$ leads to 
\begin{equation}
v_{g} = \frac{d \omega}{dk} = \frac{c - \frac{\omega}{2\sqrt{\epsilon}}~ 
\frac{\partial \epsilon}{\partial k}
}{\sqrt{\epsilon} + \frac{\omega}{2\sqrt{\epsilon}}~
\frac{\partial \epsilon}{\partial \omega}} 
= \frac{\omega}{k} 
\frac{1 - k \frac{\partial \ln n}{\partial k}  }{
1 + \omega \frac{\partial \ln n}{\partial \omega}  }  .
\label{vgroup}
\end{equation}
From Eq.~(\ref{vgroup}) it is clear that light can be slowed down
(a) due to a strong {\it frequency} dispersion $\omega d n(\omega)/ d \omega 
\gg 1$ (as discussed in \cite{einstein,eit}), 
(b) due to a small value $1-k \partial \ln n / \partial k$, i.e. when the
{\it spatial} dispersion is strong or, (c) when the wavevector $k$ becomes 
large. 
In the absence of spatial dispersion in the dielectric function the conditions 
(a) and (c) are fulfilled at frequencies corresponding to a pole in 
$\epsilon (\omega)$, where both $d n / d \omega$ and the wavevector $k$ are
large, cf. $k^2 \propto \epsilon (\omega)$.  
Furthermore, it is expected that in 
the same frequency region the dielectric function is also quite sensitive 
to the wave vector, i.e. explicit spatial dispersion is 
significant, cf. case  (b). 

Accounting for the wavevector dependence of the dielectric function, 
in general leads to multiple solutions of the dispersion relation
$c^2k^2=\omega^2\epsilon(\omega,k)$  for the wave vectors $k_{zp}$, $p=1,2$,
along the direction $z$ perpendicular to the surface at given $\omega$ in 
the geometry shown in Fig.~1(a).  
As it will be derived below, only the light-like modes with small $|k_{zp}|$ 
contribute 
significantly to the transmission and the usual one mode Fresnel result is 
recovered, if $|k_{z1}| \ll |k_{z2}|$. 
On the other hand, the conventional description breaks down, when both 
$|k_{zp}|$ are comparable and contribute to the optical properties. 
This happens if a pole in the dispersionless theory, which corresponds to
the cases (a) and (c) of low group velocity, is regularized by the introduction
of spatial dispersion.  

Depending on the type of the spatial dispersion the 
excited modes may be both real (propagating modes)
or one wave vector may be real, while the other one is complex (decaying 
mode). This leads to two types of critical frequencies, where the Fresnel 
approach becomes invalid. Namely it occurs at frequencies $\omega_e$, where 
both $k_z$ are real and $|k_{z1}|\approx |k_{z2}|$, and at frequencies 
$\omega_i$, where $k_{z1}\approx ik_{z2}$. 

\begin{figure}
\begin{center} 
\epsfig{file=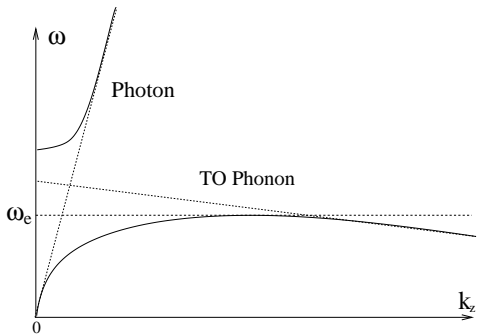,width=0.35\textwidth,angle=0,clip=}
\end{center}
\caption{Schematic mixing of a 
transverse optical phonon characterized by anomalous 
dispersion with a propagating electromagnetic wave leads to an extremal point
$\omega_e$ in the lower polariton band, where the group velocity vanishes. 
Just below the frequency $\omega_e$ two modes with similar wavevectors
propagate. 
\label{polariton} 
}
\end{figure}  

When both modes are propagating, $v_g$ vanishes at frequencies $\omega_e$
due to strong spatial dispersion, the case (b) mentioned after 
Eq.~(\ref{vgroup}), see Fig.~\ref{polariton}. 
In general, this case occurs, if the eigenmodes of the crystal, when decoupled
from electromagnetic waves,  have a dispersion opposite 
to that of the electromagnetic wave. Generic examples are 
phonon modes with anomalous (decreasing) dispersion mixing with 
propagating light of normal 
dispersion, which form a polariton (cf.~Fig.~\ref{polariton}),  
or the Josephson plasmon with normal dispersion interacting with screened 
electromagnetic waves in a superconductor, which show an 
anomalous dispersion, see Sects. III.B and IV below.  
As the main 
consequence,  near frequencies $\omega_e$ the transmission 
coefficient into the crystal is not 
determined solely by the dielectric function, but 
crucially depends on the microstructure of the crystal near the surface, if
both dissipation and disorder are very low and the system is strongly
anisotropic.  
We will also  show that interfering 
multiple propagating waves create a behavior 
similar to  intrinsic birefringence and  affect strongly the 
transmission through the crystal and multiple reflection. 

In the second situation (one mode is propagating, while another is decaying) 
the Fresnel approach breaks down near frequencies $\omega_i$, 
where the moduli of the wave vectors of two excited modes 
become equal. Near these frequencies both $|k_{zp}|$ become large, of the 
order of the inverse interatomic spacing, which leads  to a small, but finite
group velocity $v_g$ as described in the case (c) 
after Eq.~(\ref{vgroup}). This occurs, for example, 
for Josephson plasmons with anomalous dispersion in a crystal 
with different alternating junctions, where one plasmon has normal, while the 
other one has anomalous dispersion, see Sect. IV below. 
As near the frequencies $\omega_i$ only a single mode propagates into the
crystal, the transmission coefficient is significantly suppressed in
comparison with resonances at extremal points $\omega_e$, where incident 
light excites two propagating  modes.

\begin{figure}
\begin{center} 
\epsfig{file=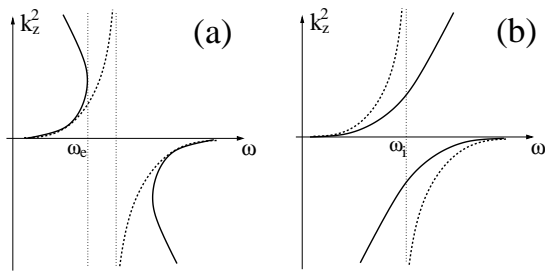,width=0.4\textwidth,angle=0,clip=}
\end{center}
\caption{
A pole in $k_z^2 (\omega)$ in the case without spatial dispersion 
(dashed line) indicates the importance of small length scales due to the 
low group velocity (cf. Eq.~(\ref{vgroup})) and the
breakdown of the macroscopic theory based on a $k_z$-independent
dielectric tensor. For an isotropic system this corresponds to a
singularity in the dielectric function, $k_z^2 \propto \epsilon (\omega)$ 
This pole is regularized, when spatial dispersion is taken into account, and 
depending on the sign of $d k_z^2 / d \omega$ an extremal point 
$\omega_e$, where
the group velocity vanishes, appears as shown in a), 
or as shown in b), the singularity transforms into a
special frequency $\omega_i$, where $k_{z1}^2 = - k_{z2}^2$. 
\label{poleschematic} 
}
\end{figure}  

In Fig.~\ref{poleschematic} it is demonstrated schematically, how the critical 
frequencies $\omega_e$ and $\omega_i$, where the amplitudes of the exited
multiple modes equal, $|k_{z1}|=|k_{z2}|$, develop from a singularity in the 
one mode theory, which  neglects the $k_z$-dependence of the eigenmodes. 
In the simplest case of an isotropic medium, which was considered after
Eq.~(\ref{vgroup}), the dispersionless dielectric function and squared
wavevector amplitudes  are proportional, $\epsilon (\omega) \propto k^2$,  
and their poles coincide.  The
breakdown of the one mode Fresnel theory at these points is already
anticipated from the 
low group velocity $v_g$, due to the large frequency dispersion,  
$|d k_z / d \omega| \gg 1$, and the large $|k_{zp}|$ near the pole, cf.
case (a) and (c) in the discussion after Eq.~(\ref{vgroup}). 

If for the crystal dispersion (without coupling to electromagnetic waves) 
$d k_z^2 / d \omega < 0 $,  an extremal point
$\omega_e$ appears below the singularity and at this frequency $\omega_e$ the
group velocity $v_{gz} = d \omega / d k_z $ vanishes 
and two propagating modes with $k_{z1} = - k_{z2}$ are excited, 
see Fig.~\ref{poleschematic}~(a). 
In a similar way, at the extremum of $\omega (k_z^2)$ above the singularity 
the imaginary excited modes merge, $k_{z1}^2 = k_{z2}^2 < 0$, while in the 
intermediate frequency region the solutions $k_{zp}^2$ are complex. 
On the other hand, if $d k_z^2 / d \omega > 0 $, the singularity in the 
dispersionless one mode theory is transformed to a special point $\omega_i$, 
where the amplitude of the excitation equals, but one is propgating and the
other decaying, $k_{z1}^2 = - k_{z2}^2$. 

Remarkably, a special point $\omega_i$ can appear, when the group velocity 
is small,  even 
without a wavevector dependence (i.e. without {\it explicit} spatial 
dispersion) 
in the dielectric function  due to the atomic structure in the unit cell alone 
({\it implicit} spatial dispersion). Generally  for each  polariton 
band a real or
imaginary mode  is excited, but usually inside one band the
second wave associated with the off-resonant excitation of the other bands can
be neglected. Here it will be shown that this assumption breaks down, when the 
group velocity becomes small, e.g. for large amplitudes of the wavevectors,
cf.  case (c). 
Thereby the system with alternating 
plasma resonances like SmLa$_{1-x}$Sr$_{x}$CuO$_{4-\delta}$ with light 
incident 
parallel to the layers (Fig.~\ref{geometry}(b) at $\theta=0$) 
presents a generic 
example, as in this case 
the wavevector $k_z$ perpendicular to the layers (explicit spatial
dispersion) vanishes due to the 
homogeneity of the incident beam. In a macroscopic theory 
the electrodynamic response to the electric field, which is averaged within
the unit cell, is decribed by the 
effective (average) dielectric function ${\tilde \epsilon}_c (\omega)$, 
\begin{equation}
\frac{1}{\tilde{\epsilon}_c(\omega)}=\frac{1}{2}
\left[ \frac{1}{\epsilon_{c1}(\omega)}+ \frac{1}{\epsilon_{c2}(\omega)}
\right].  
\label{dielecaver}
\end{equation} 
Thereby a pole  in
${\tilde{\epsilon}_c(\omega)}$ appears between the zeros of  
$\epsilon_{cl}(\omega) = \epsilon_{c0} (
1-\omega_{c0,l}^2 / \omega^2)$ ($l=1,2$, $\epsilon_{c0}$ background
dielectric constant), which correspond to the plasma 
frequencies $\omega_{c0,l}$ in the different junctions \cite{marel3,marel4}.  
This indicates the breakdown of the one-mode Fresnel
approach and the necessity to account properly for the second solution. 
Obviously, similar consequenses of such a "discrete" implicit spatial
dispersion are expected generally for any crystals with multiple optically
active crystal bands of the same symmetry.  

Both the behavior near $\omega_e$ and $\omega_i$ are in contrast to the 
conventional Fresnel theory and to the common believe that spatial 
dispersion of crystal modes or the atomic structure 
do not create measurable effects of order unity in optical
properties, but only enter in negligible corrections
proportional to the ratio of atomic scales and the wavelength of light. 
In fact,  the  Fresnel results have to be modified significantly in a 
narrow interval near the frequencies $\omega_e$ and $\omega_i$, 
but only in perfect crystals with very weak dissipation. 

Finally, we point out that the vanishing of the group velocity at extremal
frequencies $\omega_e$ due to {\it spatial} dispersion of the crystal modes 
provides a novel way to stop light pulses dynamically.  Recently it attracted 
a considerable interest to diminish the light velocity strongly with the help of 
{\it frequency} dispersive gaseous media as described by case (a) after 
Eq.~(\ref{vgroup}). 
From a practical point of view, our suggestion based on the 
${\bf k}$-dependence of the dielectric tensor allows to use 
slow light in a solid state device for the processing of information. 
In particular, the sensitivity of the group velocity in solids to the 
external fields could  be used to store quantum information in the form of 
photonic qubits, as required for optical quantum computers \cite{laflamme}. 
Our novel solid state proposal to stop light might be of advantage 
compared with the realizations using gaseous media, as it is easier to scale
 to larger system 
sizes and more complex devices. By adjusting an inhomogeneous external
 parameter, like the magnetic field for the JPR, a
spatially inhomogeneous profile for the group velocity can be imprinted.  
Such conditions can simulate in the laboratory the behavior of light 
in a curved spacetime, as realized in astrophysical situations,  
e.g. near the event horizon of a black hole \cite{leonature}.
   
Previously the spatial dispersion of the Josephson plasma mode and its 
effect on the propagating electromagnetic waves in layered superconductors
 with identical Josephson junctions was discussed 
by Tachiki, Koyama and Takahashi \cite{tach}.  
They realized that the  mixing of 
plasma modes with electromagnetic waves can lead to two propagating waves
with different wave vectors for the same frequency.  However, implications of 
this fact to the optical properties, like reflectivity, were not discussed. 
Van der Marel and Tsvetkov \cite{marel4} present an effective dielectric
function for the system with alternating Josephson junctions and charge 
coupling within the unit cell for the special case of 
incidence parallel to the layers, but they did not account correctly 
for the dissipation due to the conductivities and for the nontrivial effects 
of the ``discrete'' spatial dispersion mentioned above. 

The paper is organized as follows: In the first part, we derive in general the
optical properties of an uniaxial crystal with explicit spatial dispersion 
along
the symmetry axis in the dielectric function using additional boundary 
conditions with 
one phenomenological parameter (Sect. II).  In the second part, we  
confirm these results for oblique incidence in the microscopic model for 
the JPR accounting for the atomic (layered) structure. Thereby the ABC are 
derived and analytical solutions for 
systems with identical (Sect. III) and two different alternating 
(Sect. IV) Josephson junctions are obtained.
 In Sect. \ref{sectionparallel} the atomic structure 
is taken into account to derive the reflectivity in the incidence 
parallel to the layers. Technical details are given in the 
Appendices.

\section{Macroscopic approach for crystals with spatial dispersion} 

\label{macrosection}

In this Section we derive the dispersion relation from a macroscopic 
dielectric tensor (Sect. II.A), calculate the transmission coefficients into
(II.B) and through (II.C) the crystal using a phenomenological ABC and 
close with some further remarks, concerning e.g. future applications, like the
stopping of light (II.D).

\subsection{Dispersion relation \label{sectiondispgen}}

We consider the geometry of the incident and reflected light 
as shown in Fig.~\ref{geometry}.  
The wave vector of the incident light with frequency $\omega $ for the 
geometry shown in Fig.~\ref{geometry}(a) is ${\bf k}_0=(\omega \sin\theta
/c,0, \omega \cos\theta/ c)$, while for Fig.~\ref{geometry}(b) 
${\bf k}_0=(\omega \cos\theta /c,0, \omega\sin\theta/ c)$, where 
the $z$-axis is perpendicular to the layers (it coincides with the $c$-axis of 
the crystal).  The incident (quasi-monochromatic) electromagnetic wave 
is assumed to be P-polarized, i.e. the electric field 
${\bf E } ({\bf r} ,t)  = {\bf E} (\omega, {\bf k}) 
\exp (i {\bf k} {\bf r} - i \omega t  ) $ is in the plane 
defined by ${\bf k}_0$ and the normal of the surface ($xz$-plane), while the 
magnetic field ${\bf B}$ has only a component in $y$-direction. S-polarization 
is not considered here, as an electric field parallel to the layers does not
excite the JPR studied below.

In the macroscopic approach used here we describe the crystal by a 
dielectric tensor, which is averaged on atomic scales within the unit cell, 
but can depend on the wave vector (explicit spatial dispersion), 
and study the effects of the intrinsic microstructure (implicit spatial dispersion)
in Sect. V. 

In the following 
we will consider highly anisotropic uniaxial (layered) crystals with 
the dielectric function components $\epsilon_c(\omega,k_z)$ along the 
$c$-axis ($z$-axis) and $\epsilon_a(\omega)$ in the $ab$ ($xy$) 
plane along the layers in a parameter regime appropriate for the JPR. 
In $\epsilon_c(\omega,k_z)$ we account for a collective 
mode (JPR in our case), which  is strictly longitudinal with the 
dispersion $\omega_c(k_z)$ for $k_x=0$, i.e. 
$\epsilon_c[\omega=\omega_c(k_x=0,k_z),k_z]=0$, and whose polarization is
mainly in the $c$-direction for any $k_x$ due to the strong anisotropy, 
$|\epsilon_a|  \gg |\epsilon_c|$, near the JPR.  
We neglect the eigenmode, which is 
polarized parallel to the layers for $k_x=0$, as it is of much higher frequency 
than the JPR. 

From the bulk Maxwell equations for the Fourier components, 
\begin{eqnarray}
&& c k_x B_y = - \omega\epsilon_c (\omega,k_z)   E_z, \label{max1}\\
&& k_xE_z-k_z E_x=-(\omega/c) B_y, \label{max2} \\
&&c k_z B_y= \omega \epsilon_a (\omega)  E_x \label{max3} 
\end{eqnarray}
follows directly the dispersion relation, 
\begin{equation}
\frac{k_x^2}{\epsilon_c(\omega,k_z)}+\frac{k_z^2}{\epsilon_{a}(\omega)}=
\frac{\omega^2}{c^2}, 
\label{dr0}
\end{equation}
of the eigenmodes in the crystal. 

For the geometry shown in Fig.~\ref{geometry}(b) and neglecting the discrete 
layered structure in $z$-direction,  we obtain analogously 
 $k_z={\bf k}_{0z}=\omega \sin \theta / c$ of the excited crystal mode, while 
the dispersion relation, Eq.~(\ref{dr0}), gives a single solution for $k_x^2$.
Hence, the usual Fresnel description is generally valid, except where $|k_x|$ becomes
large, e.g. at the poles of $\epsilon_c (\omega)$, see Eq.~(\ref{dr0}). At
these  points the implicit spatial dispersion due to the atomic structure in the
unit cell in multiband systems has to be taken into account. Then 
multiple solutions  $k_x^2$ of the dispersion relation contribute, which will  be
discussed for the JPR with alternating junctions in Sect. V below.

In the geometry shown in Fig.~\ref{geometry}(a) 
we obtain from the 
translational invariance parallel to the surface the wave vector component 
$k_x={\bf k}_{0x}= \omega \sin \theta / c $, and 
the dispersion relation determines the solution(s) for the $z$-component 
$k_z(\omega,\theta)$ of the modes excited by the incident wave.
 
In a crystal described by the dielectric functions 
$\epsilon_{a,c} (\omega)$, which is independent of
the wave vector ${\bf k}$,  the dispersion relation Eq.~(\ref{dr0})
has a unique solution $k_z^2 (\omega)$.  
The Maxwell boundary conditions (MBC), 
requiring the continuity of the parallel components $E_x (z)$ and $H_y (z)$
at the surface $z=0$,
immediately give the Fresnel formula for the reflection coefficient 
$R = |r|^2$ and the transmissivity  $T=1-R$ into the crystal.  Here
\begin{equation}
r=\frac{1-\kappa}{1+\kappa}, \ \ 
\kappa=\frac{E_x (z=0) } {B_y (z=0) \cos \theta}.
\label{kappagen}
\end{equation}

When in a highly anisotropic crystal the eigenmode with electric field 
approximately parallel to the layers is neglected, 
the effective dielectric function $\epsilon_{\rm eff}$ is given by 
\begin{equation}
\kappa = \sqrt{\epsilon_{\rm eff}} = 
 \frac{n_0}{\epsilon_a(\omega)\cos\theta},
\label{kappaonemode}
\end{equation}
where the refraction index is 
\begin{equation}
n_0 = c k_z (\omega) / \omega = 
\sqrt{\epsilon_a (\omega)  (1 - \sin^2 \theta /\epsilon_c (\omega) ) }. 
\label{refractionpure}
\end{equation}
This suggests that for an anisotropic crystal in this geometry the critical 
frequencies, where the refraction index $n_0$ becomes large and the Fresnel 
theory breaks down, appear at zeros of $\epsilon_c (\omega)$ rather than at
poles of the dielectric function, as for isotropic system discussed in the 
 introduction (cf. Eq.~(\ref{vgroup})) and Fig.~\ref{geometry}(b).

If the dielectric function, $\epsilon_c (\omega, k_z)$, is dispersive 
in the $c$-direction, 
 Eq.~(\ref{dr0}) has multiple solutions for $k_z^2 (\omega)$ 
\cite{agr,tach}. In the following we restrict ourselves to the simplest case 
of four (in general complex) solutions $\pm n_1, \pm n_2$ for the refraction 
indices. 

Generally, in a crystal of finite thickness, where the (multiple) back
reflection from the second surface is taken into account, all four 
solutions $\pm n_{1,2}$ have to be considered. 
For simplicity, we will consider in the following mainly 
a semi-infinite crystal in the half space $z>0$, where 
only two of the solutions are physical. 
When dissipation is low, for quasi-monochromatic wave packets the 
direction of the energy transfer 
is determined by the Poynting vector ${\bf S}$,  which 
is oriented along the group velocity ${\bf v}_g = \partial \omega / 
\partial {\bf k}$ \cite{agr}:
\begin{eqnarray}
&&{\bf S}=W{\bf v}_g,  \label{poynting}  \\
&&W=\frac{1}{16\pi}\left[\frac{\partial(\omega\epsilon_a)}{\partial\omega}
E_x^{\phantom *} E_x^*+\frac{\partial(\omega\epsilon_c)}{\partial\omega}
E^{\phantom *}_zE_z^*+ B^{\phantom *}_yB_y^*\right]. \nonumber
\end{eqnarray}
Here $W$ is the high frequency average of the energy density.  
In  agreement with the causality principle the group velocity of propagating 
modes in the $c$-direction, 
$v_{gz}=\partial \omega(k_z)/\partial k_z$, 
 should therefore be positive. 
Note that in the case of normal (anomalous) dispersion
this requires the real part of the wave vector $k_{zp}$ (modes $p=1,2$)
and of the refraction index $n_{p} = c k_{zp} / \omega$ to be 
positive 
(negative). When dissipation is taken into account, this rule is equivalent 
to the condition that the eigenmodes should decay inside the crystal, i.e. 
${\rm Im}(k_{zp}) > 0$. 

This has in particular consequences at extremal frequencies $\omega_e$ 
of the dispersion relation ${\rm Re} ( \omega (k_z) )$, 
where  the group velocity 
$v_{gz} =0$ vanishes and two branches, one with normal  and another one 
with anomalous dispersion 
merge, see Fig.~\ref{polariton}. 
At these points the two solutions for $k_z$, which are real in the 
absence of dissipation,  have the same amplitude $|k_z|$, but different signs,
\begin{equation}
{\rm Re} [n_1(\omega_e) + n_2(\omega_e) ]  =0  .
\label{extremalpoint} 
\end{equation}

\subsection{Transmissivity $T$ on surface}

In the macroscopic approach the electric field $E_z$ and the polarization 
$P_z$ in a semi-infinite crystal with a single atom in the unit cell and 
with the background dielectric constant 
$\epsilon_{c0}$ can be expressed as
\begin{eqnarray}
&&E_z(z)=\sum_{p=1,2}E_z(k_{zp})\exp(ik_{zp}z), \\ 
&&P_z(z)=\sum_{p=1,2}E_z(k_{zp})\chi_c(k_{zp}) \exp(ik_{zp}z), 
 \\
&&4\pi\chi_c(k_z)=\epsilon_c(\omega,k_z)-\epsilon_{c0}, 
\end{eqnarray}
while the equations for $E_x$ and $B_y$, 
which enter in Eq.~(\ref{kappagen}), are similar. 
In order to determine the amplitudes 
$E_z(k_{zp})$ of the different eigenmodes we use the most general ABC 
proposed by Ginzburg \cite{agr} 
\begin{equation}
P_z(z)+\ell(\partial P_z/\partial z)=0, \ \ \ z\rightarrow 0, 
\label{ABC}
\end{equation}
where the length scale $\ell$ is a phenomenological parameter to be determined 
from the microscopic model. 
In  systems with inversion symmetry 
we can use   $\chi_c (\omega, k_z) - \chi_c (\omega, 0 ) \sim k_z^2$ for 
$k_z \rightarrow 0$ and obtain  
\begin{equation}
\sum_{p}E_z(k_{zp})(1+i\xi n_{p})=0, \ \ \ \xi=\omega\ell/c.
\label{ga}
\end{equation}
in leading order in $\epsilon_a/n_p^2 \ll 1$ and $1/|n_1 n_2| \ll 1$. In this 
limit Eq.~(\ref{ga}) and the following results are confirmed 
microscopically for the JPR in Sect.~III and IV, while in general 
corrections involving field
components parallel to the surface have to be considered in Eq.~(\ref{ABC}). 
Using Eqs.~(\ref{max3}), (\ref{dr0}) and (\ref{ga}), we derive 
(near the resonance) 
\begin{equation}
\kappa=\frac{1}{\epsilon_a \cos\theta}~\frac{
n_{1}n_{2}}{n_{1}+n_{2}-
i\xi n_{1}n_{2}}.
\label{gen}
\end{equation} 
We see that in the case of multiple eigenmodes in the crystal the optical 
properties like the reflectivity generally cannot be expressed by the refraction 
indices $n_p$ alone, which are determined by the bulk dielectric functions 
$\epsilon_{a,c}$ via Eq.~(\ref{dr0}), but also depend explicitly on the 
parameter $\xi$ introduced by the  boundary conditions. 

As the wavelength $\lambda$ of light is larger than  
all length 
scales related to the atomic structure of the crystal or to the change of the 
polarization at the surface, we can assume  $\xi \sim \ell / \lambda \ll 1$.  
Therefore the term $\xi n_1n_2$ can be neglected 
everywhere except at the extremal frequencies $\omega_e$, 
where  ${\rm Re} ( n_1+n_2 ) = 0$.

If in addition the amplitude of one excited  mode is large, i.e. 
$|n_2| \gg |n_1|$ and $|n_1 n_2| \gg |\epsilon_a|$,  the conventional one mode 
Fresnel result, Eq.~(\ref{kappaonemode}),  is obtained for the 
mode with smallest $n$.  
In Fig.~\ref{polariton} it can be seen that for the phonon
polariton away from the extremal frequency $\omega_e$ this condition is 
fulfilled
and only the usual light-like mode remains.

Deviations from the usual Fresnel theory are therefore expected, when the 
amplitudes of $n_{1}$ and $n_2$ are comparable and both modes play a role. 
The resonances in the transmissivity are located in these two mode frequency 
regions and we distinguish the cases that (i) 
both excited modes are propagating ($n_{1,2}$ real)  or (ii) one mode is 
propagating, while the second is decaying ($n_1$ real, $n_2$ imaginary). 
The appearance such type of special frequencies $\omega_e$, 
where $n_1 = - n_2$,  and $\omega_i$, where $n_1 = i n_2$,  
near a pole in the 
refraction index of the dispersionless one mode theory is schematically shown 
in Fig.~\ref{poleschematic}
 (the index of $\omega_i$ reminds of the factor $\pm i$ between the 
solutions $n_{1,2}$).

(i) For two real modes $n_{1,2}$ we have ${\rm Re} ( n_1+n_2 ) = 0$ at the 
extremal point $\omega=\omega_e$, when causality is taken into account, 
see Eq.~(\ref{extremalpoint}). 
Then, if the dissipation is weak in addition, e.g.  
${\rm Im} (n_1 + n_2) \ll | \xi  n_1 n_2| $, only the term 
$i \xi n_1n_2$ in Eq.~(\ref{gen}) remains, $\kappa(\omega_e)$ 
is imaginary and   $T (\omega_e) =0$. 
The transmissivity $T$ reaches its maximum at the frequency 
$\omega_{{\rm e, max}}$ slightly above $\omega_e$. At this frequency
\begin{eqnarray}
&&(n_1+n_2)=\epsilon_a^{-1}n_1n_2(\cos^{-2}\theta+
\xi^2\epsilon_a^2)^{1/2}, \label{omegamax} \\
&&T_{\rm e,max}=2/[(1+\xi^2\epsilon_a^2\cos^2\theta)^{1/2}+1]. 
\label{Tmax1macro}
\end{eqnarray} 
It is pointed out that both the position $\omega_{\rm e,max}$ of the resonance 
in $R$ or $T$ and its amplitude are determined not solely by the imaginary
part of $\epsilon_{a,c}$ as in the dispersionless case,  
but also by the surface parameter $\xi$. This correction is important for
highly anisotropic systems, where $\xi \epsilon_a \gg 1$, although $\xi \ll
1$, as it is realized for the JPR (see Eq.~(\ref{xiepsa})). 
We see that in the absence of dissipation 
$T_{\rm e,max}$ depends on $\xi$ and is generally 
smaller than the Fresnel result $T_{\rm max} =  1$, see 
Fig.~\ref{RTschematic}. 
Physically this result reflects the fact that the low 
group velocity near $\omega_e$ introduces a small length 
scale $\lambda_g = v_{g} / \omega$, which makes  
the variation of the polarization $P_z$ near the surface relevant and indicates  
the breakdown of the translational invariance on the atomic scale $l$. 
Note that the opposite signs  of the refraction indices 
$n_{1,2}$ near $\omega_e$ due to causality are essential for the dependence 
of $T_{\rm e,max}$ on $\xi$. 
The vanishing of $n_1 + n_2$ at $\omega_e$ (see Eq.~(\ref{extremalpoint})) 
in Eq.~(\ref{gen}) and its 
consequences in Eqs.~(\ref{omegamax}) and (\ref{Tmax1macro}) 
have not been noted previously \cite{pekar,pekar2,agr,henn,chen,forst}.  
We also note that the results in Eqs.~(\ref{omegamax}) and 
(\ref{Tmax1macro}) cannot be 
obtained from the ABC proposed by Pekar \cite{pekar,pekar2}, which neglects 
the derivative in Eq.~(\ref{ABC}).

(ii) In the case,  when $n_1$ 
is real, while $n_2$ is imaginary without dissipation, we anticipate that 
$T$ is strongly suppressed, because both modes are excited by the incident 
light, but 
only  a single mode propagates into the crystal. This situation occurs e.g. 
in superconductors when the dispersion of the collective mode is anomalous 
(cf. Fig.~\ref{schematicmixing2layer} in 
Section \ref{sectionalternate}). $T(\omega)$ is in this case peaked at 
critical 
frequencies $\omega_i$ near $\omega_c$, where $n_2 =- i n_1$ with $n_1 < 0$.
Here we obtain for the maximal transmission coefficient 
\begin{equation}
T_{{\rm i,max}} = T (\omega_i ) = \frac{2 n_1 }{\epsilon_a \cos \theta},  
\label{tmaxi}
\end{equation}
so that $T(\omega_i)\ll T_{{\rm e,max}}$. 
This difference in the resonance amplitude, depending whether two or one
propagating modes are excited, cannot be described in the one mode Fresnel 
approach without spatial dispersion, where in both cases a single propagating 
mode is excited and the transmission amplitudes are comparable. 
This observation and the strong deviation from the conventional Fresnel 
result is confirmed below for the JPR in Sect. IV. 
In contrast to the situation (i) near extremal points $\omega_e$, the 
parameter $\xi$ is irrelevant near $\omega_i$.

\subsection{Transmission through thin film}

\begin{figure}
\begin{center} 
\epsfig{file=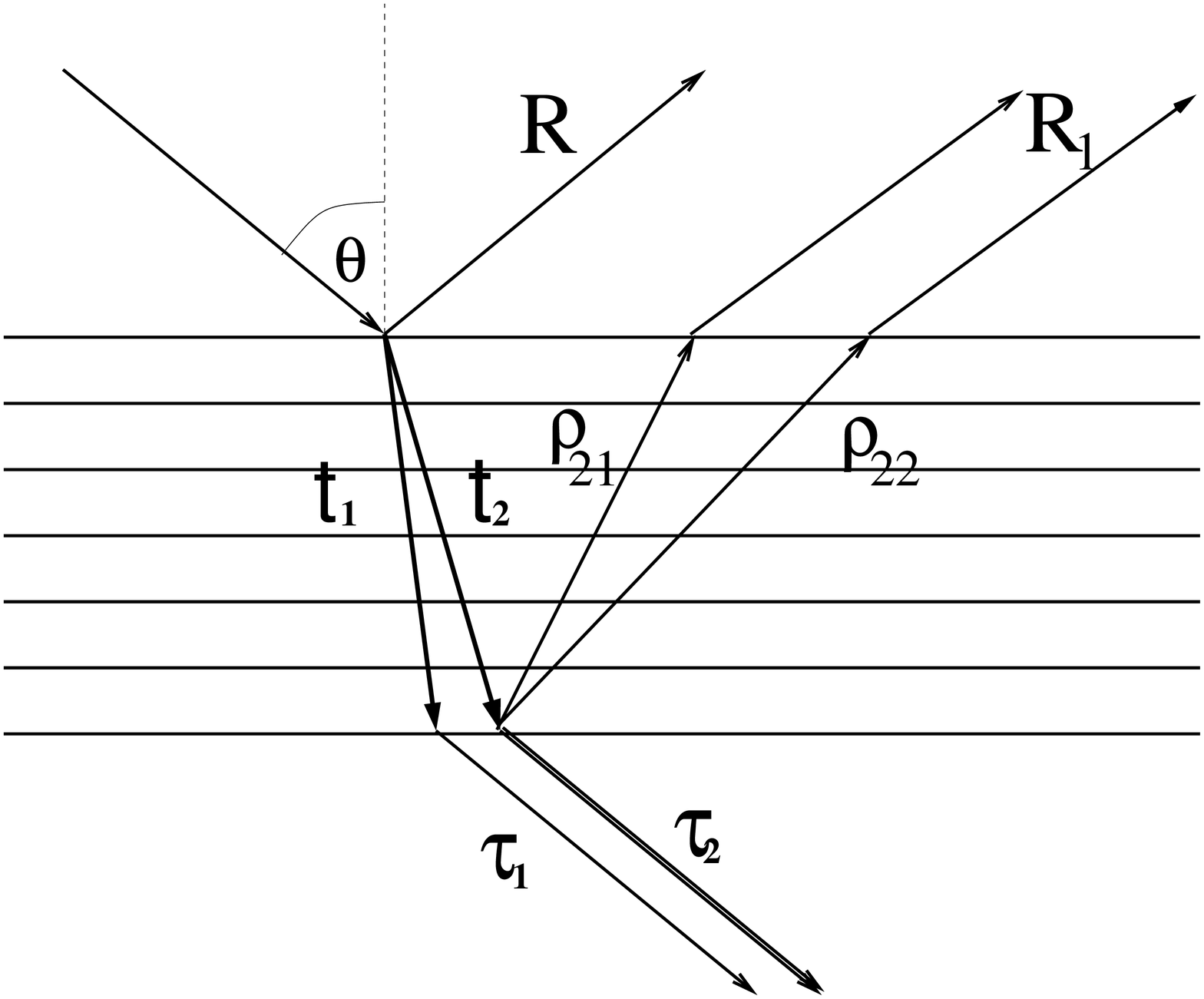,width=0.3\textwidth,angle=0,clip=}
\end{center}
\caption{
Transmission amplitudes of the wave with refraction index $n_p$ 
into ($t_p$) and out of ($\tau_p$) the crystal and 
multiple reflections at the first ($R_n$) and the second surface 
($\rho_{pp^\prime}$).
\label{multirefl} 
}
\end{figure}

We now study the transmission and back reflection of the multiple excited 
modes in a thin film of finite thickness $L$, see Fig.~\ref{multirefl}. 
For the ratio of the magnetic field $t_p B_{y}^{\rm in}$ of a partial wave
with the refraction index $n_p$ ($p=1,2$) excited in the crystal to that of 
the incident wave $B_y^{\rm in}$ we obtain 
\begin{equation}
t_p=i(-1)^p\frac{2(1-i\xi n_p)}{\xi(n_2-n_1)(1+\kappa)}.
\end{equation} 
We will see that $|t_p|>1$ for JPR, e.g. the fields of the two partial 
waves are enhanced, 
but have opposite direction. 
Note that the transmissivity $T$ follows from the ratios of the $z$-components of 
the Poynting vectors, Eq.~(\ref{poynting}), and that $T \neq |t_1+t_2|^2$. 

\begin{figure}
\begin{center} 
\epsfig{file=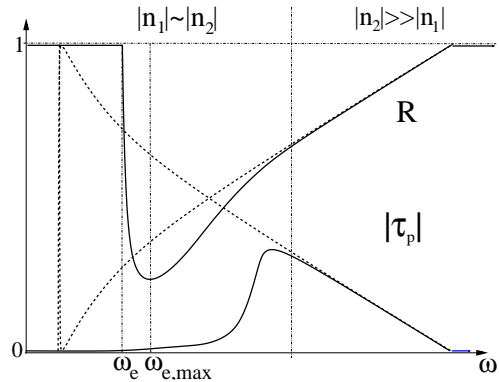,width=0.36\textwidth,angle=0,clip=}
\end{center}
\caption{Reflectivity $R=1-T$ and ratio $|\tau_p|$ of the outgoing 
 magnetic fields at the second surface of the crystal near an extremal point 
$\omega_e$  with (solid line) and 
without (dashed) spatial dispersion without dissipation (schematically). 
Compared with the conventional Fresnel formulas the plasma edge in $R$ is at
the higher frequency $\omega_{e}$ 
and the amplitude of the resonance at $\omega_{\rm e,max}$ 
is damped due to $\xi$ in Eq.~(\ref{gen}). The amplitude  $\sim 
|\tau_p(\omega_{\rm e,max})|$ of the outgoing waves (cf. Eq.~(\ref{taul}))
is strongly 
suppressed in the frequency region, where R is minimal, e.g. where the 
transmission $T$ into the crystal is maximal (Eq.~(\ref{taumax})).  
\label{RTschematic} 
}
\end{figure}  

At the second surface of a crystal 
 the arriving wave with index $n_p$ ($p=1,2$)
and the magnetic field amplitude ${\tilde B}_{y,p}$
creates a wave, which is emitted out of the crystal.  Its wave vector is 
${\bf k}_0$ and we denote its magnetic field  by $\tau_p {\tilde B}_{y,p}$. 
Each wave $n_p$ also excites two waves with refraction indices $n_{p^\prime}$ 
and magnetic fields $\rho_{pp^\prime} {\tilde B}_{y,p}$, 
which are reflected back into the crystal. 
The ABC Eq.~(\ref{ABC}) at $z=L$ for these three waves give
\begin{equation}
(1+i\xi n_p) {\tilde E}_{z,p} + (1-i \xi n_1) {\tilde E}_
{z,p1}^{\rm ref} + ( 1 - i \xi n_2 ) {\tilde E}_{z,p2}^{\rm ref} = 0, 
\end{equation}
where ${\tilde E}_{z,p}$ and ${\tilde E}_{z,p p^\prime}^{\rm ref}$ 
are electric field components
at the second surface at $z=L$ of the arriving and back-reflected waves, 
respectively. We find in leading order in $\xi$ ($\bar{p} =3-p$)
\begin{eqnarray}
&& \tau_p=\frac{2 n_p(n_1+n_2)}{
(n_1+n_2)\epsilon_a\cos\theta+\epsilon_a+n_1n_2}, 
\label{taul} \\
&& \rho_{pp} = (-1)^p \frac{(n_1+n_2)}{(n_2-n_1)} 
\frac{\epsilon_a \cos \theta (n_2-n_1) + \epsilon_a - n_1 n_2}{
\epsilon_a \cos \theta (n_2+n_1) + \epsilon_a + n_1 n_2 }  , \\ 
&& \rho_{p \bar{p}} = \frac{(-1)^p 2 n_p (n_p^2  - \epsilon_a ) 
}{ (n_2 - n_1) ( \epsilon_a \cos \theta (n_1 + n_2) + \epsilon_a + 
n_1 n_2) }   . \label{rho2}
\end{eqnarray}
At the frequency $\omega_{\rm e,max}$, where the transmissivity $T$ into the 
crystal is maximal, the transmission, 
\begin{equation}
\tau_p ( \omega_{\rm e,max} ) = \frac{2 n_p}{\epsilon_a \cos \theta },
\label{taumax}
\end{equation} 
is strongly suppressed in comparison with the conventional Fresnel result 
(cf.~Fig.~\ref{RTschematic}). 
At the same point $\omega_{\rm e,max}$ the back scattering takes place almost 
completely into the same eigenmode, $\rho_{11} \approx - \rho_{22}
 \approx - 1 + O (n /\epsilon_a )$ and $\rho_{12} = 
\rho_{21} \approx 1 /[ \epsilon_a \xi 
\cos \theta] \ll 1$, while at $\omega = \omega_e$ we obtain 
$|\rho_{pp}| \gg \rho_{12}
\rho_{21}$ in the presence of spatial dispersion. 

The two eigenmodes of the same polarization interfere inside the crystal 
and for the total transmission $T_{\rm tot} = |t_{\rm tot}|^2$
through the sample we obtain near an extremal frequency $\omega_e$
\begin{eqnarray}
t_{\rm tot}&=&\sum_l t_l\tau_l\exp(in_l\omega L/c) \\ 
   &\sim&  [1+(1-2v)\cos(2n\omega L/c)]/2 \, , 
\label{interfere}
\end{eqnarray} 
where $v\approx (n \xi / 2)^2$. Therefore, the transmission coefficient has 
oscillatory behavior as a function of the frequency $\omega$ and the sample 
thickness $L$ due to the
interference effect, even if the back reflection into the sample
is irrelevant. 
Near the frequency $\omega_{\rm e,max}$ multiple reflection leads to 
\begin{equation}
T_{\rm tot} = | t_{\rm tot} |^2 \sim \frac{1+(1-2v)\cos(2n\omega L/c)}{
      1 + \rho^2 - 2 \rho  \cos(2n\omega L/c) }
\end{equation}
with $\rho = \rho_{12} \rho_{21}$. 

The difference with conventional birefringence lies in the fact 
that all waves have the same P-polarization.  This type of so-called intrinsic 
birefringence has also been observed in semiconductors for certain directions
of propagation (cf. \cite{burnett} and references therein), 
while in the present case it appears for an arbitrary angle of (oblique) 
incidence. 
Alternatively, the effect of spatial dispersion can be observed by the 
splitting of a spatially focused incoming beam into two outgoing ones, 
corresponding to the two different group velocities in the crystal
(angle between rays $\sim 10^{-3}$ degrees for JPR).

\subsection{General remarks}

Some additional remarks to the macroscopic approach are in place: 

(1) It is pointed out that even if the last term $\sim \xi$ in the 
denominator in 
Eq.~(\ref{gen}) can be neglected for frequencies far from the band edge near
$\omega_{{\rm e,max}}$ or due to dominant dissipation, the interplay of the 
two modes with indices $n_{1,2}$ can lead to unconventional effects, like 
intrinsic birefringence (Eq.~(\ref{interfere})) or the suppression of the
 transmission near $\omega_i$ in comparison with the Fresnel result 
(Eq.~(\ref{tmaxi})).  
Only in the limit $|n_2|\gg |n_1|$ and $|n_1 n_2| \gg |\epsilon_a |$ the smallest 
refraction index determines $\kappa$, $\tau_p$ and $\rho_{pp^\prime}$ and 
the usual one-mode Fresnel description is recovered.  

(2) Thereby the existence of a pole in the effective dielectric function in 
the one-mode Fresnel approach is an indication of the existence of 
a special point $\omega_e$ or $\omega_i$, see Fig.~\ref{poleschematic} and 
the microcopic confirmation in the Sects. \ref{sectionalternate} 
and \ref{sectionparallel}.  However, we point out that without further 
investigation of the spatial dispersion or the atomic structure 
these two cases cannot be distinguished.The guiding picture in 
Fig.~\ref{poleschematic} and the microscopic results for the JPR in oblique 
incidence in Sect. III and IV  and for phonon polaritons \cite{ivopaper} 
suggest that special points of type $\omega_e$ ($\omega_i$) appear, if light
is  mixed with a crystal mode of opposite (same) dispersion. This 
is seen in Fig.~\ref{schematicmixing2layer}, where the mixing of the plasma
 band in the lower 
(upper) band with normal (anomalous) dispersion with decaying light creates 
a special point of type $\omega_e$ ($\omega_i$).  
In Sect. V it is shown that special frequencies $\omega_i$, where 
$n_1^2 = - n_2^2$, can appear near the pole of the 
effective dielectric function even without ${\bf k}$-dependence due to the 
discrete atomic structure within the unit cell. 

(3) It is stressed that the Kramers-Kronig relations  
expressing causality (and sum rules following from them) 
are still valid in the two-mode regime for 
physical response functions like the reflectivity $R$ or for the effective
dielectric function $\epsilon_{\rm eff} = \kappa^2$ extracted from $R$, but do
not apply to the refraction indices $n_p$ of the partial waves independently 
\cite{agr,kramerskronig,sumrules}. 

(4) We note that beyond the universal electrodynamic effects studied above 
there
might also be the necessity that the ABC reflect the change of the internal 
structure of the crystal excitations near the surface. This problem has 
been studied in detail for the  Frenkel exciton, which is quite extended on  
the atomic scale and whose wave function is consequently modified near the 
surface, see Ref.~\onlinecite{pekar2,agr,birman1,birman2,chen,forst} and 
references therein. 
Due to the  focus on  the  microscopic derivation of the exciton modes
and despite a considerable effort, some of the crucial general features 
discussed here have been missed for that system, namely the importance of 
the atomic structure 
(parameter $\xi$) and  the correct causal choice of the eigenmodes 
in a semi-infinite crystal near the extremal points, 
e.g. $n_1 + n_2 \approx 0$ for $\omega \approx \omega_e$, see 
Refs.~\onlinecite{agr,birman1,birman2,chen}. 

In the case of the JPR the 
effect of the surface on the internal structure of excitations turns out 
to be very weak,  because the excitations are confined between layers on 
the atomic scale and in 
highly anisotropic layered superconductors the layers near the surface are 
practically the same as those inside the crystal.  
Therefore and because we discuss this system only as a generic example for 
general electrodynamic features, which are relevant for a large class of 
systems, we will not address this question in the following and assume a 
dielectric response function 
$\epsilon_c(z,z^\prime)= \Theta (z) \Theta(z^\prime) \epsilon ( z-z^\prime )$. 

(5) The dispersion and the group velocity of phonon polaritons has been 
measured directly by exciting locally 
a wave packet and detecting the time of propagation to a separated probe 
position in the crystal \cite{timeofflight}. Future experiments of this type
with high resolution for long wavelengths could also show the existence of
extremal frequencies $\omega_e$, where the group velocity $v_{gz}$ vanishes 
at a finite wave vector as shown in Fig. \ref{polariton}.

(6) We now comment on the perspectives  to stop light using  
{\it spatial} dispersion at extremal frequencies $\omega_e$  
(cf. Fig.~\ref{polariton}) and compare this method with 
the alternative one, which uses the {\it frequency} dispersion of the 
dielectric function \cite{eit}. 

The effect of the frequency and/or spatial dispersion on the group velocity 
has already been discussed as a guiding principle for an isotropic medium, see 
Eq.~(\ref{vgroup}). In the scattering problem depicted in 
Fig.~\ref{geometry}~(b) the component $k_x$ of the wavevector and the group
velocity parallel to the layers is fixed by the 
boundary condition. The signal velocity $v_{gz}$ in $z$-direction  
in the anisotropic case ($n_{a} = \sqrt{\epsilon_{a}}  \neq n_c = 
\sqrt{\epsilon_c}$) follows from Eq.~(\ref{dr0}), 
\begin{equation}
v_{gz} = \frac{d \omega}{d k_z} = \frac{\omega}{k_z} 
\frac{
 n_c^2 k_z^2 ( 1 - \frac{\partial \ln n_a}{\partial \ln k_z} ) 
- \frac{\partial \ln n_c}{\partial \ln k_z} n_{a}^2 k_x^2  
}{
n_c^2 k_z^2 ( 1 + \frac{\partial \ln n_a}{\partial \ln \omega} ) + 
 \frac{\partial \ln n_c}{\partial \ln\omega} n_{a}^2 k_x^2
} . 
\label{vgroupaniso}
\end{equation}

In the phenomenon of electromagnetically induced transparency (EIT), which has 
recently been used to create ultra-slow light \cite{eit}, atomic levels are 
pumped optically in such a way that the medium exhibits a sharp 
absorption line in ${\rm Im} (\epsilon (\omega))$
 near a resonance frequency $\omega_0$
for propagating light. According to the 
Kramers-Kronig relation the frequency dispersion $d n / d \omega$ of the real
part of $\epsilon$ is therefore quite large, which suppresses the group 
velocity in Eq.~(\ref{vgroup}). Spatial dispersion $\omega(k)$ is discussed 
here for the first time as a tool to stop light, although a finite drift 
velocity of a (gaseous) medium has been interpreted in this way
\cite{drift}. 

This effect  might be used to realize certain phenomena connected with
ultra-slow light in a solid, such as the optical Aharonov-Bohm effect in
rotating  media \cite{leoslow1} or the enhanced two-photon interaction via 
a phonon mode \cite{2photon}, which has possible applications in quantum
information processing. 

Apart from this, the variation of the band structure and  thus
$\epsilon_{a,c} ({\bf r})$ on scales, which are large compared with the 
wavelength $\lambda$ of light, allows to manipulate the geometrical optics of 
light in a solid in a rather simple way, e.g. via a space dependent external
magnetic field for the JPR or pressure for phonon modes.  
Similar feature have been used recently   for creating artificially
 local space-time geometries, which are reminiscent of cosmological 
phenomena, such as black holes: e.g. in superfluid $^3$He \cite{volovik}, 
 inhomogenously pumped media with EIT \cite{leonature}, flowing dielectrics
\cite{schuetzhold} or solids \cite{reznik}. In particular, it is possible to 
create a space dependent group velocity
profile for a given frequency, where $v_{gz}$
vanishes on some manifold in space. At this point the behavior of light is
expected to be  similar to the one near an event horizon of a black hole, see
\cite{leonature}. Thereby the description in terms of a dielectric function 
breaks down at short length scales.

From an application point of view, the modification of the band structure with 
the help of an external parameter, opens 
the perspective to store light pulses dynamically. 
Thereby in an ideal crystal the phase information of the light pulse or the 
single photon is stored coherently, which makes the device potentially useful
in quantum information processing \cite{laflamme}. 
The limiting factor is clearly the
decoherence due to disorder or dissipation induced by a finite conductivity.
For the JPR in Bi$_2$Sr$_2$CaCu$_2$O$_8$  the intrinsic decay
time due to ohmic losses is estimated as $\tau \sim 10^{-8} s \sim 10^5 
\tau_{\rm osc}$, while the oscillation frequency 
$\tau_{\rm osc} \sim 10^{-12} s$ is in the THz regime. 
Although an adiabatic switching of the external magnetic 
field appears necessary, 
a certain number of quantum manipulations seems to be possible.  
While in metals or semiconductors the decoherence will be prohibitively high, 
defect free insulators might be much better than this estimate. 
On the other hand, a solid state realization of a memory unit for a quantum
computer has obvious advantages in terms of scalability to devices of higher
complexity in comparison with EIT based systems.

(7) While on the one hand the above results are applicable to a 
wide variety of 
systems, strictly speaking the use of the ABC Eq.~(\ref{ABC}) 
can only be justified in a microscopic model, where also the 
parameter $\xi$ has to be determined.  
This will be accomplished in the following for the JPR, because 
there the problem can be formulated as a set of linear finite difference 
equations and therefore a complete solution for all wave vectors can be 
obtained. 

Thereby it turns out that the optical properties of crystals with several 
atoms in the unit cell cannot be described by the function $\epsilon_c 
(\omega, k_z)$
alone. Then the above macroscopic approach based on the slowly 
varying polarization $P_z({\bf k})$, which is reflected in the ABC 
Eq.~(\ref{ABC}), breaks down for both oblique incidence and incidence parallel 
to the layers, see below Sect. IV and V. 

\section{Microscopic approach for JPR in crystals with identical junctions} 

\subsection{General equations}

Considering a stack of identical  Josephson junctions, we label the layers 
by the index $m$, the interlayer spacing is 
$s$ and the intrinsic Josephson junctions are characterized by the critical 
current density $J_{0}$.  
Thus the plasma frequency at zero wave vector is given as 
\begin{equation}
\omega_{c0}^2=\frac{8\pi^2csJ_0}{\epsilon_{c0}\Phi_0}=\frac{c^2}
{\lambda_c^2\epsilon_{c0}}, 
\label{pl}
\end{equation}
where $\Phi_0$ is the flux 
quantum and $\lambda_c$ is the penetration length along the $c$-axis 
\cite{kosh,lnb,bul}.

In order to determine the transmissivity in the microscopic approach, 
we solve the 
Maxwell equations inside the crystal by accounting for 
 supercurrents inside the
2D layers at $z=ms$ and interlayer Josephson and quasiparticle currents, 
which are driven by the difference $V_{m,m+1}$ 
of the electrochemical potentials in neighboring layers: 
\begin{eqnarray}
&&c\frac{\partial B_y}{\partial z}=
i\epsilon_{a0}\omega\left[E_x-\frac{\omega_{a0}^2}
{\omega^2}\sum_{m=0}^{N}E_xs \delta (z-ms)\right], \label{first} \\
&&\frac{\partial E_x}{\partial z}-ik_xE_z=i\frac{\omega}{c} B_y, \ \ 
E_{z,m,m+1}= \int \limits_{ms}^{(m+1)s}E_z\frac{dz}{s},  \label{second} \\
&&c k_xB_y=-\omega\epsilon_{c0}\left[
E_z-\sum_{m=0}^NP_mf_{m}(z)\right], \label{e} \\ 
&& \frac{{\tilde \omega}^2 es}{\omega_{c0}^2} P_m= V_{m,m+1} 
=esE_{z,m,m+1}+\mu_{m+1}-\mu_{m}. \label{last}
\end{eqnarray}
Thereby $\omega_{a0}=c/\lambda_{ab}\sqrt{\epsilon_{a0}}$ is the in-plane
plasma frequency, $\epsilon_{a0}$ is the high frequency in-plane 
dielectric constant and the function $f$ is defined as
$f_{m}(z)=1$ at $ms<z<(m+1)s$ and zero outside this interval. 
It is seen from Eq.~(\ref{e}) that the discrete quantity 
$P_m=(1/s) \int_{ms}^{(m+1)s} P_z (z) dz$ 
plays the role of the $z$-axis polarization $P_z (z)$
averaged between the layers $m$ and $m+1$, as it 
describes the  response of the Josephson plasma oscillations to the electric 
field in junction $m$. For small amplitude
oscillations the supercurrent density is given by the phase difference 
$\varphi_{m,m+1}=2ie V_{m,m+1}/\hbar\omega$ as
$J_{m,m+1}^{(s)}=J_0\sin\varphi_{m,m+1}\approx J_0\varphi_{m,m+1}$, 
which was used to derive Eq.~(\ref{e}). 
The difference $\mu_m-\mu_{m+1}=(4\pi 
s\alpha/\epsilon_{c0})({\rho}_m-{\rho}_{m+1})$,  of the chemical potentials
$\mu_m$ can be expressed by the 2D charge densities, ${\rho}_m$, which in 
turn are related to the electric fields $E_z (z=ms \pm 0)$ near the layers
by the Poisson equation, $4 \pi \rho_m = E_z (ms+0) - 
E_z (ms-0) $.  
Further, $\tilde{\omega}^2=\omega^2(1-i4\pi\sigma_c \omega / \omega_{c0}^2
\epsilon_{c0}^{\phantom 2})^{-1}$  contains the dissipation due to 
quasiparticle tunneling currents, $J_{m,m+1}^{(qp)}=\sigma_c V_{m,m+1}/es$, 
which are determined by the conductivity $\sigma_c$ and driven by the 
difference $V_{m,m+1}$ of the {\it electrochemical} potentials. Note that the
assumption in \cite{koy} that the quasiparticle current is driven by the 
averaged electric field $E_{z,m,m+1}$ is an inconsistent treatment of
the dissipation \cite{sendai}.

For 2D free electrons 
we get $\partial \mu/\partial {\rho}=\pi\hbar^2/(e m_e)$ and we can estimate
the order of $\alpha=( \epsilon_{c0} / 4 \pi e s ) (\partial \mu / \partial 
\rho)$ as $\approx 0.38$, assuming 
 $s=6.3$ ${\rm \AA}$ and $\epsilon_{c0}=20$.  This agrees well with 
$\alpha \approx 0.4$, which was  extracted in the one-layer compound  
SmLa$_{1-x}$Sr$_{x}$CuO$_{4-\delta}$ from the magnetic field 
dependence of the plasma peaks in the loss function in parallel incidence
both in the liquid \cite{ourpaper} and the solid phase \cite{newpaper}. 
The apparent free electron value of the electronic
compressibility of the CuO$_2$-layers is not in a contradiction to 
the slightly 
enhanced effective mass $m^*$ seen in ARPES measurements \cite{norman}, 
as both quantities are renormalized differently by interactions.
For systems with 
CuO$_2$ multilayers smaller values for the compressibility are anticipated
due the enhanced density of states, effective mass $m^*$, lattice constant
$s$ and the smaller background dielectric constant $\epsilon_{c0}$, 
namely $\alpha \sim 0.05 - 0.1$ for Bi-2212 or Tl-2212
(assuming $\epsilon_{c0} \approx 10$ and $d \approx 12 \AA$), but
this quantity can only be extracted reliably from experiment.  
The modification of the dispersion due to nonequilibrium effects
is not considered in the following, e.g. it is assumed 
that all frequencies are smaller than the charge imbalance and energy 
relaxation rates \cite{art,sendai,ryndykneu}. 

\subsection{Dispersion relation \label{disp1layersec} } 

We obtain now the dispersion relation for eigenmodes inside the bulk
crystal.  For an infinite number of junctions
we average Eqs.~(\ref{first})~-~(\ref{last})  between the layers $m$ and 
$m+1$ and neglect the discrete layered structure, when treating the derivatives
with respect to $z$ in the Eqs.~(\ref{first}) and (\ref{second}), 
i.e. we replace $E_{x} (z=ms)$ by $ E_{x,m,m+1} = \int_{ms}^{(m+1)s} dz E_x$  
and $B_y (z=ms)$ by $B_{y,m-1,m} = \int_{(m-1)s}^{(m)s} dz B_y$. 
Using the Fourier representation with respect to the discrete 
variable  $m$ this gives Eq.~(\ref{dr0}) with 
\begin{eqnarray}
&&\epsilon_c(\omega,q)=
\epsilon_{c0}[1-\omega_c^2(q)/\tilde{\omega}^2], \label{maxwellaverages} \\
&&\omega_c^2(q)=\omega_{c0}^2[1+2\alpha(1-\cos q)], \nonumber \\ 
&&  \epsilon_{a}(\omega)=\epsilon_{a0}(1-\omega_{a0}^2/\omega^2)
\end{eqnarray}
where $0\leq q\leq 2\pi$ and $\omega_c^2(q)$ 
describes the dispersion of the plasma mode propagating along the $c$-axis. 
Using  Eq.~(\ref{dr0}) with $k_z^2=2(1-\cos q)/s^2$, which reflects the 
existence of an upper edge of the plasma band, we obtain the dispersion of 
eigenmodes  propagating inside the crystal in an arbitrary direction. 
Due to  $\omega^2\epsilon_{a}(\omega)\approx 
- c^2/\lambda_{ab}^2$ at $\omega \approx \omega_{c0} \ll \omega_{a0}$ 
we get in the absence of dissipation ($\sigma_c=0$):
\begin{eqnarray}
\frac{\omega^2(k_x,q)}{\omega_{c0}^2} &=&
1+ \frac{\lambda_c^2k_x^2}{1+(2\lambda_{ab}^2/s^2)(1-\cos q)} \nonumber  \\ 
&&+2\alpha(1-\cos q).
\label{dd1}
\end{eqnarray}
The first term in 
the right hand side of Eq.~(\ref{dd1}) is due to the inductive coupling of 
the in-plane currents excited by the component $E_x$ of the electric field.
 The second term reflects the $c$-axis dispersion due to the  charge coupling 
of the intrinsic junctions, which is mediated by variations of the 
electrochemical potential on the layers.  
For $\alpha=0$ this dispersion of the plasma mode 
has already been calculated in Ref.~\onlinecite{lnb}. 


For the geometry shown in Fig.~\ref{geometry}(a) we can express
$k_x=\omega\sin\theta/c$ via the frequency $\omega$
and the angle $\theta$ of the incident wave and obtain the dispersion 
relation for 
the eigenmodes, which are excited by external electromagnetic waves, 
\begin{equation}
w = \frac{\omega^2}{\omega_{c0}^2}= 
 1+2\alpha(1-\cos q)+\frac{(a-1)\beta}{\beta+1-\cos q}  . 
\label{d1}
\end{equation}
Here  $\beta=s^2/(2\lambda_{ab}^2a) \sim 10^{-4}$ describes the 
inductive coupling 
and $a^{-1}= 1- c^2 k_x^2 / (\omega^2 \epsilon_{c0}) =
1-\sin^2\theta/\epsilon_{c0}$.  To include
dissipation, one has to replace $\omega$ and $w = \omega^2 /\omega_{c0}^2$ by 
${\tilde \omega}$ and ${\tilde w}$ in Eqs.~(\ref{dd1}) and (\ref{d1}). 

\begin{figure}
\begin{center}
\epsfig{file=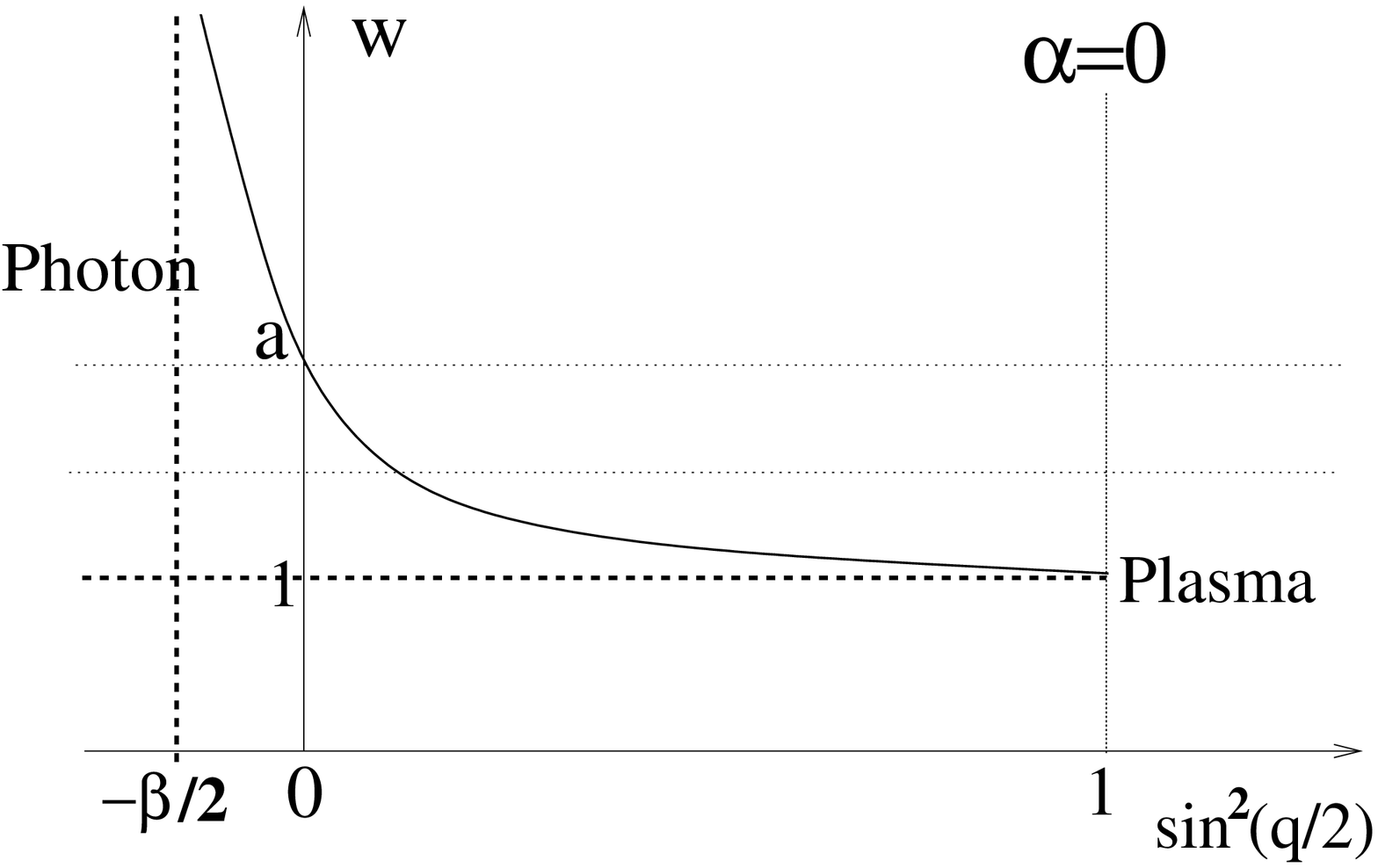,width=0.4\textwidth,angle=0,clip=}
\newline
\epsfig{file=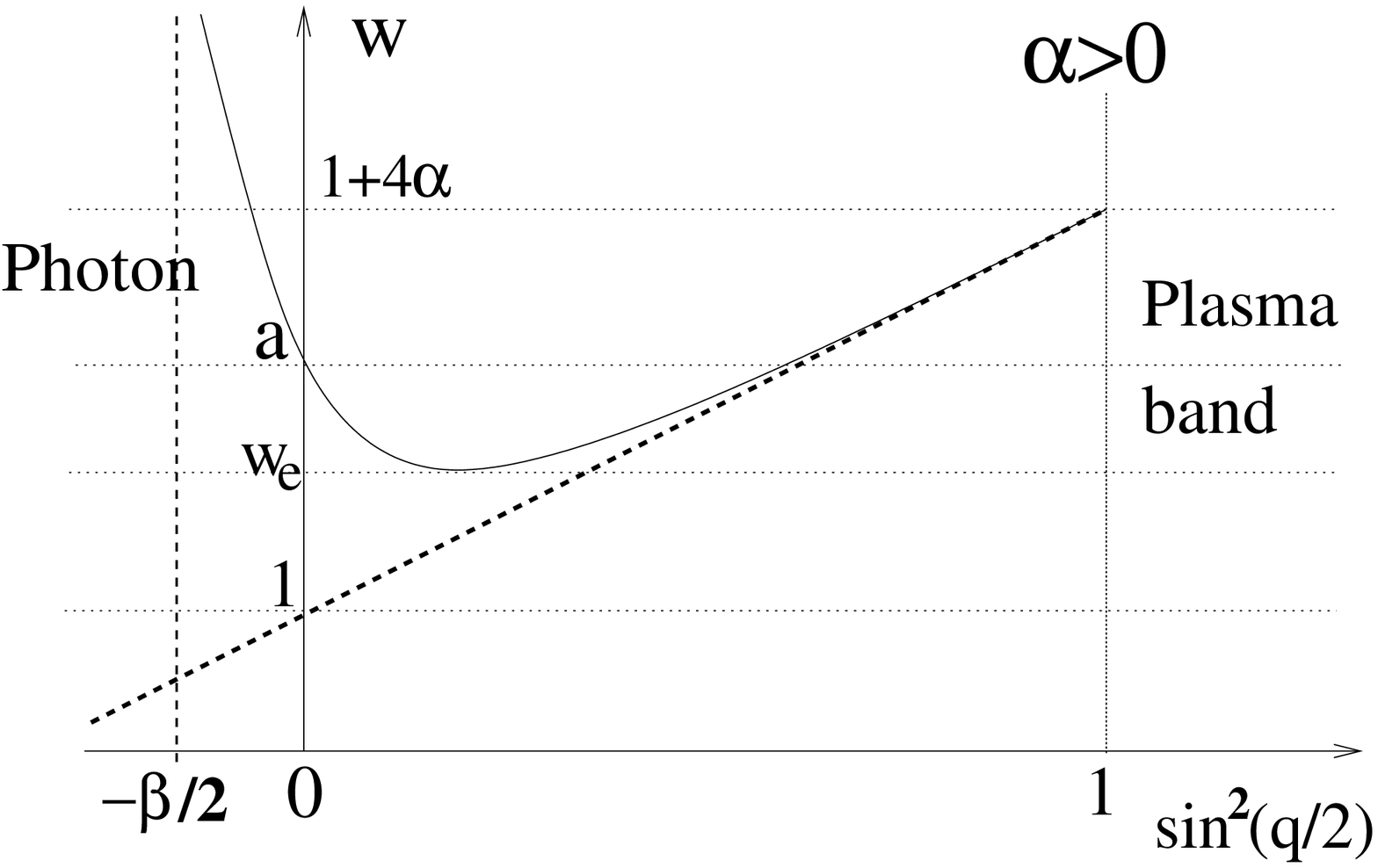,width=0.4\textwidth,angle=0,clip=}
\end{center}
\caption{
Schematic picture of the dispersion relation, 
$w = \omega^2 / \omega_{c0}^2$, depending on 
$\nu^2=\sin^2(q/2)$ (solid line), Eq.~(\ref{d1}), for $\alpha=0$ (above) and 
$\alpha \neq 0$ (below).  
$0<\nu^2<1$ corresponds to propagating solutions
 with real $q$, while outside this interval $q$ is complex and the modes 
decay. 
It is seen that the mixing of a decaying electromagnetic wave 
(dashed line at $\nu^2 = - \beta /2$) with the plasma 
band with normal dispersion $\alpha \neq 0$ (dashed)  leads to an extremal point 
$w_{\rm e}$ and a region $w_{\rm e} < w < a (\theta)$, 
where two propagating eigenmodes with normal and anomalous dispersion exist. 
\label{schematicmixing1layer} 
 }
\end{figure}

In Fig.~\ref{schematicmixing1layer} we plot schematically the dispersion 
$w = \omega^2 /\omega_{c0}^2$ 
versus $\nu^2=\sin^2(q/2)$. Thereby, $\nu$ is a 
normalized form of the refraction index $n=(2c/\omega s)\nu$ and can be 
used to present both propagating ($q$ real, $\nu^2  \in [0,1]$)
and decaying (${\rm Im} (q) \neq 0$, $\nu^2  \notin [0,1]$) modes.

In the absence of charge coupling, $\alpha=0$, the eigenmode, which is excited
in oblique incidence ($a (\theta) \neq 1$), has anomalous dispersion, 
$\partial \omega(\nu)/\partial \nu<0$, cf. Fig.~\ref{schematicmixing1layer}
above. 
It is seen that at $\alpha=0$ the width $a-1$ of the transmission
window $w \in [1,a]$,
 where modes can propagate into the crystal, are determined by the
extremal values at $\sin(q/2) = 0$ and $\sin (q/2) = 1$. 

For normal incidence $\theta=0$ ($\Leftrightarrow a=1$) 
the longitudinal plasma mode with $\alpha \neq 0$ is decoupled from the \
transverse 
electromagnetic wave as shown by the dashed lines in 
Fig.~\ref{schematicmixing1layer} below,
 because the electromagnetic wave does not 
have an $E_z$ component which excites plasma oscillations between the layers.  
In this case the wave vector of the pure electromagnetic wave inside
 the crystal is given by the 
relation $1-\cos q+\beta=0$, i.e. the electromagnetic wave decays 
on the scale $\lambda_{ab}$ due to the screening in the conducting layers.  
On the other hand, the wave vector of the propagating longitudinal plasma 
mode, $q$, is given by the relation $w=1+2\alpha(1-\cos q)$ and it is 
real in the frequency interval $\omega_{c0}\leq\omega\leq\omega_{c0}
(1+4\alpha)^{1/2}$. The pure plasma mode has a normal dispersion, 
$\partial\omega(\nu)/\partial \nu\geq 0$.

As $a > 1$ ($\Leftrightarrow \theta \neq 0$)
is close to unity for any angle $\theta$ and $\epsilon_{c0}\approx 10$,
the parameter $\beta = s^2/ 2 a \lambda_{ab}^2 \sim 10^{-4} \ll 1$
is  small and the two modes mix only when the second and third term 
in Eq.~(\ref{d1}) are approximately equal. This happens 
at small $\nu^2 = \sin^2(q/2)
\approx u / 8 \alpha$, where the small scale $u$ is given as 
$u = [8 (a-1) \beta \alpha]^{1/2}$ ($u \sim 10^{-3}$ for cuprates). 
For any angle $\theta \neq 0$ the modes inside the crystal are a mixture 
of the longitudinal plasma oscillation and the transverse electromagnetic 
waves. As a consequence, the electric and magnetic fields of the eigenmode 
are not polarized parallel or perpendicular to the wave vector, 
i.e. the eigenmodes are neither purely transverse nor longitudinal.

From Fig.~\ref{schematicmixing1layer} below it is clear 
that the mixing of these two 
degrees of freedom at $a\neq 1$ and nonzero $\alpha$ can lead to the 
existence of an extremal point $w_e$, where the character of the dispersion 
changes and the group velocity vanishes. This happens at $\omega_e=1+u$, 
provided that $\alpha > (a-1) \beta /8$  and the dissipation is weak,
i.e. ${\rm Im} (n_p) \ll {\rm Re} (n_p) $
or equivalently  
${\tilde \sigma} = 4 \pi \sigma_c / \omega_{c0} \epsilon_{c0}  \ll u$.
We estimate ${\tilde \sigma} \sim u$ in Bi-2212 \cite{latyshev}, 
${\tilde \sigma} > u$ in SmLa$_{1-x}$Sr$_{x}$CuO$_{4-\delta}$ 
\cite{ourpaper,newpaper}
or other cuprates with $d$-wave order parameter.  Layered s-wave superconductors 
with the JPR frequency in the optical interval would be perfect candidates to  
study  the effects of spatial dispersion, because their quasiparticle 
conductivity is very low at low temperatures (such systems are possibly
realized in organic superconductors \cite{organic} or 
intercalated LaSe(NbSe$_2$) \cite{nbse1}, which has a large anisotropy 
$B_{c2,ab} / B_{c2,c} \sim 50 - 130$ and is therefore expected to be
a Josephson coupled system \cite{nbse2}). 

In coincidence with the general picture presented in Fig.~\ref{poleschematic}
the extremal point $\omega_e$ appears near the plasma frequency 
($w=1$),  where the wavevector in $c$-direction in the dispersionless theory 
gets large, see Fig.~\ref{schematicmixing1layer}(a). 
This point corresponds to a zero in the dielectric function $\epsilon_c
(\omega)$,  as expected from the one mode Fresnel theory, cf. 
Eq.~(\ref{refractionpure}).

\begin{figure}
\epsfig{file=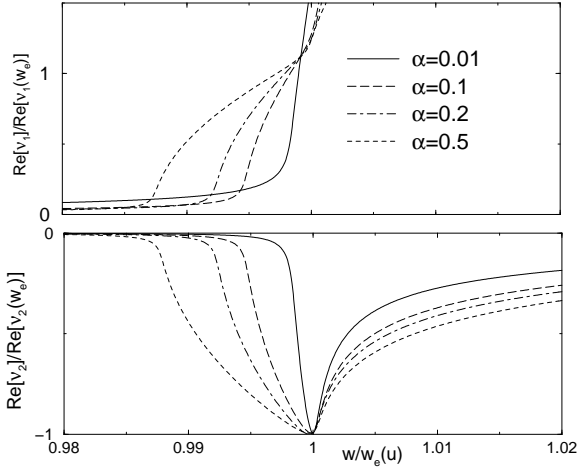,width=0.35\textwidth,angle=-90,
clip=}
\caption{Real part 
of $\nu_1$ (above) and $\nu_2$ (below) as a function of the 
normalized squared frequency $w/w_e$
near the plasma resonance for different 
$\alpha$ (${\tilde \sigma}=4 \pi \sigma / \epsilon_0 \omega_{c0,1} = 0.26$, 
$\beta=10^{-4}$, $a(\theta) =1.1$).  
For $w_e<w<a$ causality requires that ${\rm Re} (\nu_1) > 0 $ 
(${\rm Re} (\nu_2)<0 $) for the 
solutions with normal (anomalous) dispersion. 
In the interval $1- u < w < w_{\rm e}$ we have 
in particular  ${\rm Re}(\nu_1) \approx - {\rm Re} (\nu_2)$, e.g 
standing waves due to interference of $\nu_1$ and $\nu_2$, 
see Eqs.~(\ref{inter1}) and (\ref{inter2}).   
\label{Ren1layer}
 }
\end{figure}

\begin{figure}
\epsfig{file=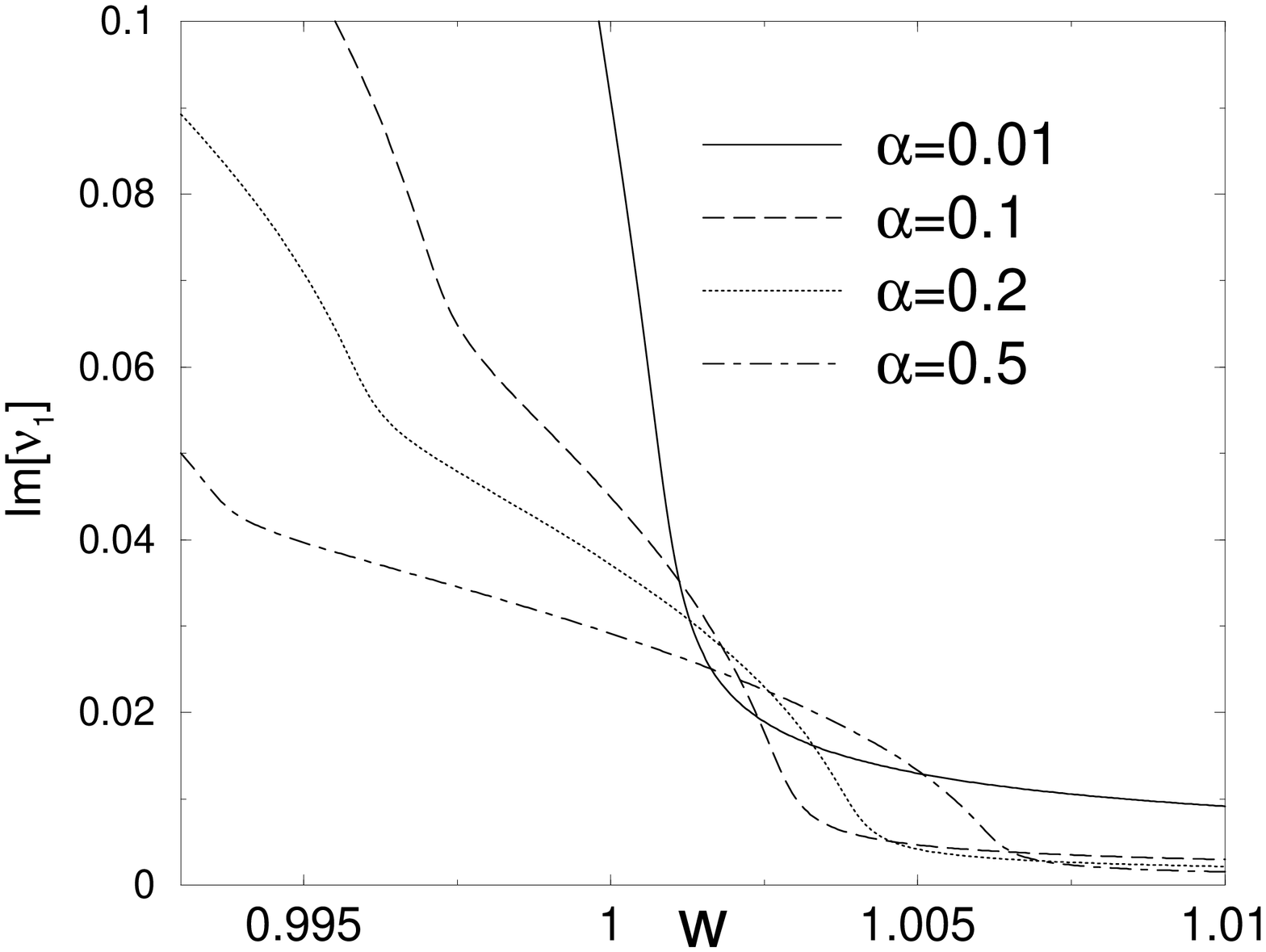,width=0.35\textwidth,angle=-90,clip=}
\newline
\epsfig{file=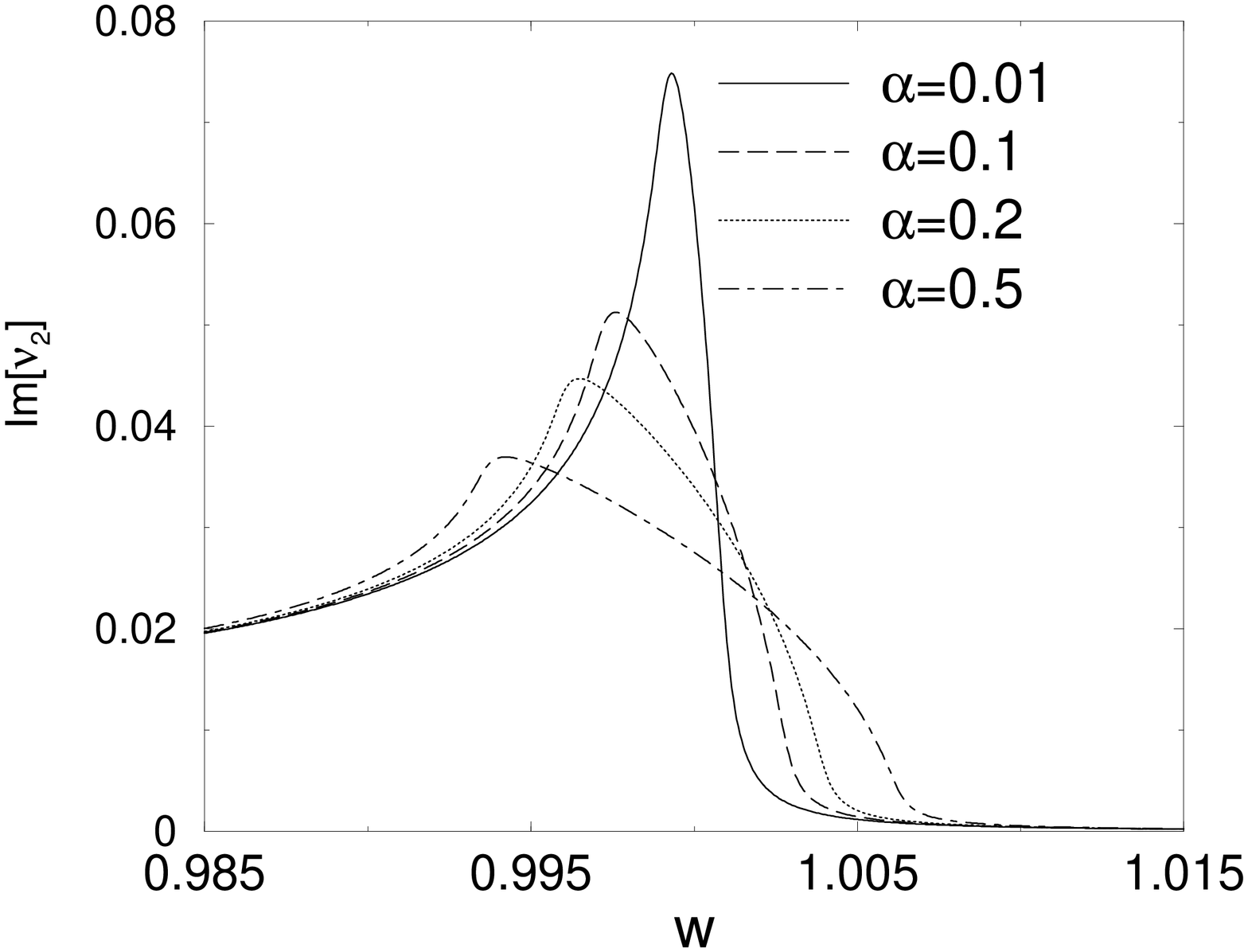,width=0.35\textwidth,angle=-90,clip=}
\caption{ Imaginary part of $\nu_{1,2}$ for different values of $\alpha$
(${\tilde \sigma} =0.26$, $\beta=10^{-4}$, 
$a=1.1$). In the region $1-u<w<w_e$ below the plasma band we have 
 ${\rm Im}(\nu_1) \approx {\rm Im} (\nu_2)$. 
\label{Imn1layer}
}
\end{figure}

In the general case of nonzero dissipation
Eq.~(\ref{d1}) has four complex solutions for $\nu_{1,2}$ at 
given $\tilde{w}=\tilde{\omega}^2/\omega_{c0}^2$,
\begin{eqnarray}
\nu_{1,2}^2(\omega)&=&(\tilde{w}-1-2\alpha\beta)/8\alpha \pm \label{q} \\
&&[(\tilde{w}-1-2\alpha\beta)^2+8\alpha\beta(\tilde{w}-a)]^{1/2}/8\alpha . 
\nonumber
\end{eqnarray}
Near the lower band edge ($\omega \approx \omega_e$) this simplifies to 
\begin{equation}
\nu_{1,2}^2(\omega)= [{\tilde w} -1 \pm \sqrt{({\tilde w}-1)^2 - u^2 }  ] / 8
\alpha \; . 
\end{equation}
Therefore we obtain 
\begin{equation}
n^2= |n_1 n_2| =  \lambda_c^2\epsilon_{c0}u/2\alpha s^2
\sim \lambda_c^2/(s\lambda_{ab})\gg 1
\label{nsquestimate}
\end{equation}
 in the case of JPR. 

As discussed in Sect.~\ref{sectiondispgen}, in a semi-infinite crystal 
only those modes are physical, which decay inside the crystal, i.e. 
${\rm Im} (\nu)>0$, see Fig. \ref{Imn1layer}. 
For propagating modes this implies that the group velocity  obeys causality,
$v_{gz} >0$, and ${\rm Re} (\nu_1) >0$ (${\rm Re} (\nu_2)<0$) for 
branches with normal (anomalous) dispersion, see Fig.~\ref{Ren1layer}.

We discuss first the limiting case with vanishing dissipation 
($\sigma_c \rightarrow 0$), where 
the solutions inside the crystal are either  exponentially decaying 
($q$ imaginary) or propagating modes ($q$ real). 
For $\alpha=0$ we obtain propagating mode with real $q$  
in the frequency range $1\leq w\leq a$  (cf. dispersion in 
Fig.~\ref{schematicmixing1layer}), 
and exactly in this interval the reflection coefficient $R<1$.  
For finite $\alpha \neq 0$ two physical solutions with real $q$ exist 
in the interval $w_{\rm e} = 1+u \leq w \leq a$ provided that
$\alpha>(a-1)\beta/8$. In the range $a<w \leq 1+4\alpha$ one 
wave vector, $q_1$, is real while the other, $i|q_2|$, is imaginary.  
The important point is that this  evanescent solution has small 
$|q_2|\leq 2\beta \ll 1$ and because of this it affects 
strongly the optical properties, which are sensitive to large length scales. 
Outside of the interval $[w_{e},1+4\alpha]$ both $q_p (\omega)$ are
imaginary. 

While in the absence of dissipation within the plasma band  
$w_e<w<1+4 \alpha$ 
at least one of the eigenmodes propagates into the crystal, 
for $w \ll 1$ we obtain 
$\nu_{1,2}^2 < 0 $ and the modes $q_1$  and $q_2$ decay
rapidly on the scales $\sqrt{\alpha} s$  and $\lambda_{ab}$ respectively. 

In the intermediate regime, $1-u<w<w_e$, we have  
\begin{eqnarray}
&&{\rm Re}(q_1)= - {\rm Re} (q_2) = 
 [(u+w-1)/4\alpha]^{1/2},  \label{inter1}  \\
&&{\rm Im}(q_1)= {\rm Im} (q_2)   = [(u-w+1)/4\alpha]^{1/2},   \label{inter2}
\end{eqnarray}
and the real and imaginary parts of the wave vector $q$ are of the same order
$\sqrt{u}$ (cf.~Figs.~\ref{Ren1layer} and \ref{Imn1layer}). 
Therefore, they penetrate deep into the crystal 
and form standing waves, which decay  and oscillate on the scale 
$[2 \lambda_{ab}s\sqrt{\epsilon_{c0} a} / \sin \theta]^{1/2}$. 
 In fact, they are intermediate between modes at $w \ll 1-u $, 
which decay much faster, and propagating modes at $1+4\alpha>w>w_e$.
  
\subsection{ Eigenmodes of a semi-infinite crystal}

The averaged Maxwell equations (\ref{max1})-(\ref{max3}) are sufficient to 
determine the bulk dispersion relation Eq.~(\ref{dr0})
of the excited eigenmodes and to identify possible critical frequencies 
$\omega_e$ or $\omega_i$, where the amplitudes of the excited modes equal. 
At these points the group velocity is expected to be low and the microscopic 
layered structure  has to be considered more accurately in order to describe 
optical properties.

For this purpose we solve the electrodynamic equations between the layers 
$m$ and $m+1$ 
by using Eqs.~(\ref{first})~-~(\ref{last}), namely the equation 
\begin{equation}
g^{-2}\frac{\partial^2B_y}{\partial z^2}+B_y=a\sin\theta P_m, \ \ 
g=\frac{\omega}{c}~\left(\frac{\epsilon_{a0}}{a}\right)^{1/2}.
\label{exmode}
\end{equation}
Physically Eq.~(\ref{exmode}) describes the excitation of a propagating 
intrajunction
mode with the polarization of the electric field in $x$-direction. Thus at 
$ms\leq z\leq (m+1)s$ the solutions for the fields are
\begin{eqnarray}
&&B_y(z)=C_m\exp(igz)+D_m\exp(-igz)+a\sin\theta P_m, \label{def1} \\
&&E_x(z)=(\epsilon_{a0}a)^{-1/2}[C_m\exp(igz)-D_m\exp(-igz)], 
\nonumber \\
&&E_z(z)=(\sin\theta/\epsilon_{c0})[C_m\exp(igz)+D_m\exp(-igz)]+aP_m.\nonumber
\end{eqnarray}
Directly from the Maxwell equations   follow the continuity relations, 
\begin{eqnarray}
 E_x (z = m s +0   )  &=& E_x (z = m s - 0 )  , \label{bc1}  \\ 
 B_y (z = m s +0 ) &=& B_y (z = m s - 0 ) + 4 \pi s J_{x,m} ,   \label{bc2}  
\end{eqnarray} 
for the fields $B_y$ and $E_x$ at layer $m$ with a parallel current
$4 \pi J_{x,m} = i \omega_a^2 E_x (z=ms) / \omega $. 
Together with Eqs.~(\ref{def1}) this leads to 
the following set of equations for $c_m=C_m\exp(igd(m+1/2))
\sin\theta /\epsilon_{c0}$ and $d_m=D_m \exp(-igd(m+1/2)) \sin\theta/\epsilon_{c0}$
inside the crystal ($N-2\geq m \geq 1$, $N$ is the number of junctions):
\begin{eqnarray}
&&c_m\eta^{-1}-d_m\eta-c_{m-1}\eta+d_{m-1}\eta^{-1}=0, \label{e1} \\
&&2(c_m\eta^{-1}-c_{m-1}\eta)+(a-1)(P_m-P_{m-1})+ \nonumber \\ 
&&i(\beta/b)(c_m\eta^{-1}-d_m\eta)=0, 
\label{e2} \\
&&P_m({\tilde w}-a)+\alpha(P_{m+1}+P_{m-1}-2P_m)= \nonumber \\
&&(\sin (b) / b)(1-2\alpha \beta)(c_m+d_m), \label{e3}
\end{eqnarray}
where $\eta=\exp(ib)$ and the small parameter $b=gs/2\sim s/\lambda_c  
\sim 10^{-5} \ll 1$ 
characterizes the discreteness of the crystal structure.   
We will assume in the following 
that $\beta,b\ll\sqrt{\beta}$ and $q \sim \beta^{1/2} \sim 2/ \lambda_{ab} $, as it 
is fulfilled for highly anisotropic ($\lambda_c \gg \lambda_{ab}$) layered 
superconductors, e.g. Bi- or Tl-based cuprates.  In our 
calculations we will keep only the terms of 
lowest order in the small parameters $\beta$ and $b$.  
Eqs.~(\ref{e1}) - (\ref{e3}) give the 
dispersion relation Eq.~(\ref{d1}) with high accuracy $b^2$ and 
$\alpha \beta$. 
This difference between the exact result following from the 
Eqs.~(\ref{e1}) to (\ref{e3}) and the averaged dispersion 
(cf. Eqs.~(\ref{maxwellaverages})) can be understood explicitly from 
 Eqs.~(\ref{def1}): the replacement of $E_x (ms) $ by the 
averaged $E_{x,m,m+1}$  is correct in order $b$., i.e. when neglecting 
the discrete layered structure within the unit cell. 

The solution inside the crystal has the form 
\begin{eqnarray}
&&c_m=\sum_{p=1,2}\gamma_p\exp(iq_pm), \label{eqc}  \\
&&d_m=\sum_{p=1,2}\gamma_p d(q_p)\exp(iq_pm),
 \\
&&P_m=\sum_{p=1,2}\gamma_p P (q_p)\exp(iq_pm), 
\label{eqA} \\
&&d(q)=\frac{1-\eta^2\exp(-iq)}{\eta^2-\exp(-iq)},  
\label{dq} \\
&& P(q)=\frac{1+d(q)}{w-a-2\alpha(1-\cos q)},
\label{aq}
\end{eqnarray}
where $q_p(\omega)$ are the wave vectors of the eigenmodes for a given frequency 
as determined by  Eq.~(\ref{q}) and $\gamma_p$ denote the relative amplitude
of the excited modes, which is to be determined next. 

Neglecting the layered structure, e.g. $b \sim s/\lambda_c \rightarrow 0$ and 
$\eta \rightarrow 1$, we obtain $c=d=1$. In this case we can relate the
variables $c_m$, $d_m$ and $P_m$ with the electric and magnetic field averaged 
between the layers, i.e.  $E_{x,m,m+1} \propto c_m + d_m$  and $E_{z,m,m+1}$
is mainly determined by the polarisation $P_m$. 

\subsection{Microscopic Boundary condition}

Now we find the ratio of the 
amplitudes  $\gamma_1$ and $\gamma_2$ microscopically
by solving the electrodynamics of the surface junctions explicitly rather 
than using any phenomenological ABC. 
The equations for the first superconducting layer ($m=0$), which are 
complementary to the Eqs.~(\ref{e1})-(\ref{e3}),   read  as
\begin{eqnarray}
&&c_0+d_0 +(a-1)P_0= \frac{\sin\theta}{\epsilon_{c0}}(B_{y}^{\rm in}
+B_{y}^{\rm ref}), 
\label{eq1}\\
&&\frac{c_0\eta^{-1}-d_0\eta }{\sqrt{\epsilon_{a0}a}}=
\frac{\sin 2\theta}{2\epsilon_{c0}}(B_{y}^{\rm in}-B_{y}^{\rm ref}),
 \label{eq2} \\
&&P_0({\tilde w} -a)+\alpha(P_1-P_0-aP_0)-  \nonumber \\
&&(1+\alpha)(c_0+d_0)=\alpha\sin\theta(B_{y}^{\rm in}
+B_{y}^{\rm ref}).  \label{eq3}
\end{eqnarray}
Here $B_{y}^{\rm in},B_{y}^{\rm ref}$ 
are the magnetic fields for incident and reflected light 
respectively.  We omitted in these equations terms proportional to 
$\beta/\sin(q/2)$ and  
$b/\sin(q/2)$, which are of order $\beta^{3/4} \sim \epsilon_a / n_p^2 \sim 
(s/\lambda_{ab})^{3/2} \ll 1$ and 
$b/\beta^{1/4} \sim 1/ n^2 \sim s \lambda_{ab} / \lambda_c^2 \ll 1 $ 
in comparison with remaining terms of order unity.  
After eliminating the fields $B_{yi}$ and $B_{yr}$ from Eqs.~(\ref{eq1}) - 
(\ref{eq3}) we obtain in lowest order in $b$ and  $\beta$
the microscopic boundary condition 
\begin{eqnarray}
&&P_0({\tilde w} -a)+\alpha ( P_1-2P_0) - \label{mbc} \\
&&\alpha(\epsilon_{c0}+1)[c_0+d_0+(a-1)P_0]-  
(c_0+d_0)=0.
\nonumber 
\end{eqnarray}
We can present this condition 
in a more transparent form by calculating the difference between 
Eq.~(\ref{e3}) for $m=0$, 
where $P_{-1}$ (outside the crystal) is
formally given by Eq.~(\ref{eqA}) for $m=-1$, and the real equation for 
$P_0$, Eq.~(\ref{mbc}).  
We also take into account that in the lowest order in 
$\beta$ and $b$ we obtain
the relation $c_0+d_0+aP_0\approx P_0$ near $w_e$ with accuracy 
$\beta^{1/2} \sim (\epsilon_a/n_p^2)^{2/3} \ll 1$ using Eqs.~(\ref{dq}) and 
(\ref{aq}).  This gives the boundary condition 
\begin{equation}
P_{m=-1} =   \sum_{p=1,2} \gamma_p  P (q_p)\exp(-iq_p) = 0 , 
\label{MBC}
\end{equation} 
which has the simple interpretation that the surface 
junction ($m=0$) has only one neighboring junction, i.e. the junction 
$m=-1$ is absent. 
This result is a  microscopic derivation of the ABC Eq.~(\ref{ABC}) by noting 
that $P_m$ is the average macroscopic polarization $P_z(z)$ 
between neighboring layers, i.e. 
\begin{equation}
P_{-1} = \frac{1}{s} 
\int_{-1}^0 P_z(z) \approx P_z (z=0) - s \partial_z P_z (z=0) . 
\end{equation}
Taking into account that the deviation of $R$ from unity is significant only 
when $|q_p| \ll 1$, 
we expand Eq.~(\ref{MBC}) in $q_p$ by using $P(q) - P(0) 
\sim q^2$ 
(in leading order in $b$) from Eq.~(\ref{aq})  and obtain  
Eq.~(\ref{ga}) with $\ell=-s$: 
\begin{equation}
\sum_{p=1,2} \gamma_p ( 1 - i q_p ) = \sum_{p=1,2} \gamma_p (1 + i \xi n_p  )
=0 . 
\end{equation}
Note that this result and consequently also the expression for $\kappa$, 
Eq.~(\ref{gen}), is only valid in leading order in $\epsilon_a / n_p^2 
\sim \beta^{3/4}  \ll 1$. 

With this identification of the parameter $\xi$ we can estimate  
\begin{equation}
\xi |n_1 n_2| \sim \lambda_c/\lambda_{ab}\gg 
\xi \epsilon_a  \sim s \lambda_c / \lambda_{ab}^2 > 1
\label{xiepsa}
\end{equation}
at $\omega=\omega_e$ in Bi- and Tl-based layered superconductors. This shows
that when the anisotropy $\lambda_c / \lambda_{ab} $ is large enough, the
atomic structure modifies strongly the transmission, cf. 
Eqs.~(\ref{omegamax}) and (\ref{Tmax1macro}). 
Here we also justify the relations discussed in Sect. II, 
\begin{equation}
\xi |n_1 n_2| \sim \frac{\lambda_c}{\lambda_{ab}} \ll n_1 + n_2 \sim 
\frac{\lambda_c}{(s \lambda_{ab})^{1/2}}, 
\end{equation}
which  allow us to neglect the atomic structure away from a small frequency 
intervall of  width $\sim u^{1/2}$ around $\omega_e$.

Due to $|n_1| \gg |n_2|$ (cf. Fig.~\ref{Ren1layer}) and 
$|n_1 n_2| \epsilon_a^{-1} \sim \lambda_{ab} / s \gg 1$ away from $\omega_e$
or $\omega_i$ the usual one-mode Fresnel theory is valid everywhere, except 
near the resonances at $\omega_{e,i}$.

\subsection{The transmission coefficient}

As a consequence, we reproduce in our microscopic theory 
Eq.~(\ref{gen}) 
for $\kappa$ and therefore the transmission and reflection coefficients 
$T$, $\tau_p$, $\rho_{pp^\prime}$, Eqs.~(\ref{Tmax1macro}) and 
(\ref{taul})~-~(\ref{rho2}).

\begin{figure}
\epsfig{file=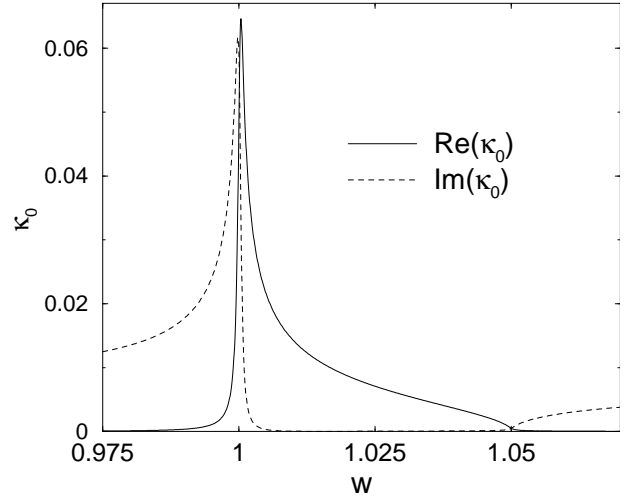,width=0.38\textwidth,angle=-90,clip=}
\caption{
The dependence of the real and imaginary part of 
$\kappa_0 = \kappa (\cos\theta \epsilon_a \xi / 2 ) $ on 
$w= \omega^2 / \omega_{c0}^2$ for ${\tilde \sigma} = 5 \; 10^{-4}$ and 
$\alpha=0.001$. The lineshape of $\kappa_0$ 
is asymmetric with a sharp edge at the extremal frequency $w_{\rm e} = 1+ u 
\approx 1$ and the upper edge at $w=a(\theta)=1.05$, which is determined by
the angle $\theta$ of incidence.  
\label{kappa1layer}
}
\end{figure}

\begin{figure}
\epsfig{file=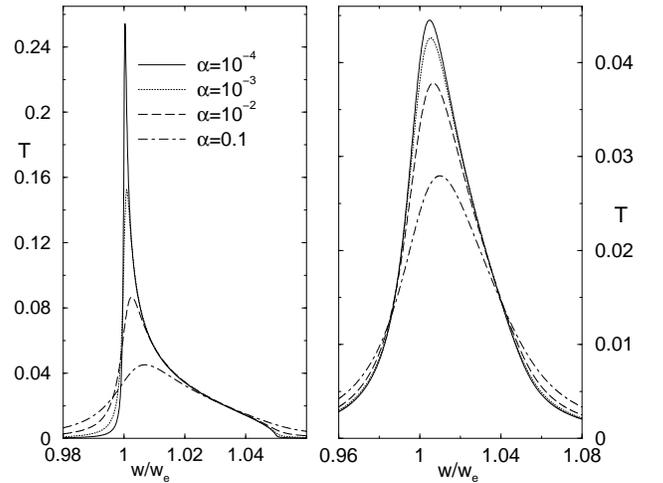,width=0.36\textwidth,angle=-90,clip=}
\caption{
Transmission $T$  depending on $w/w_e$ near the JPR frequency for 
conductivities ${\tilde \sigma} =10^{-7} $ (left) or  ${\tilde \sigma} = 0.01$
(right) for various $\alpha$ ($\beta=10^{-4}$, $a=1.1$, $\xi
\epsilon_a \cos\theta =2 $). For low dissipation ${\tilde \sigma} \ll u^2$
(left) the resonance is additionally damped due to $\alpha$ 
in the region near $w_e$, where the Fresnel approach is invalid. 
\label{kappaalpha1layer}
}
\end{figure}

The real and imaginary parts of $\kappa = \epsilon_{\rm eff}^2$ are shown in 
Fig.~\ref{kappa1layer} and have a characteristic shape with a  sharp edge 
at the extremal point 
$w_e$, provided that the dissipation ${\tilde \sigma}$ is small. 
The real part is dominant only in the interval $w_e < w < a$, 
where both modes $\nu_1$ and $\nu_2$ are real and propagating and 
the transmission $T$ into the crystal is significant. The window of
transmission $w_{\rm e,max} \approx w_{e} \leq w \leq a $ 
is therefore only determined by $a(\theta)$, and
not by the bandwidth $\sim \alpha$. The width of the peak
in $\kappa$ and $T$ near $w_{\rm e,max}$ assuming a 10\% criterion
is of the order $100u$.

In the interval $a<w<1+4\alpha$, where $\nu_1 \sim i \beta^{1/2}$ becomes 
imaginary and small,  while $\nu_2$ is real, 
$\kappa$ is a complex number (even for 
$\sigma_c=0$) with a real part proportional
to $|\nu_1^2|$.  In contrast to the standard Fresnel 
expressions, this makes transmission possible, but it is weak of the order 
$b$, because  only a small part 
of the incident light transforms into a  propagating mode.  
Therefore, deviations of $R$ from unity are significant only in the frequency 
range $ w_e \approx 1\leq w\leq a$, as in the system without dispersion. 

If the dissipation is very weak,
\begin{equation}
{\rm Im}(n_1 + n_2) \ll \xi n_1 n_2  
\Leftrightarrow
{\tilde \sigma} \ll u^2,
\end{equation}
the nonuniversal term characterized by the parameter $\xi$ in 
Eq.~\ref{gen} is important. Then according to 
Eq.~(\ref{omegamax}) the maximum of $T$ is reached at 
$w_{\rm e,max}=w_e+u^{3/2} / \sqrt{8\alpha}$.  The amplitude, 
\begin{equation}
T_{\rm e,max}=\frac{2}
{[1+ (s\lambda_c\cos\theta/(\sqrt{\epsilon_{c0}}\lambda_{ab}^2))^2]^{1/2}+1}, 
\end{equation}
is smaller than unity and it depends on the microscopic structure via 
the 
factor $s\lambda_c\sqrt{\epsilon_{c0}}/\lambda_{ab}^2$ which may be of 
order unity in 
cuprates like Tl-2212 with $\lambda_c/\lambda_{ab}\sim 100$ 
and the JPR frequency $\sim 20$ cm$^{-1}$.  
This effect can be seen in Fig.~\ref{kappaalpha1layer} (left): 
Without dispersion, i.e. ${\tilde \sigma} \gg u(\alpha)^2$ for 
$\alpha = 10^{-4}$, 
the peak amplitude is limited by the small dissipation, ${\tilde \sigma}$, 
only, while for $\alpha =0.1$ (${\tilde \sigma} \ll u^2$) the peak at 
$\omega_{\rm e,max}$ is damped {\em additionally} due to the novel term 
$i \xi n_1 n_2$ in Eq.~(\ref{gen}), as discussed above.
Physically this can be understood from 
the fact that the vanishing group velocity leads to a slow motion of the 
wave-packet and hence makes the transmission sensitive to the inhomogeneous 
layered structure of the system, i.e.  the translational invariance of
the system is broken. 

On the other hand, high dissipation ${\tilde \sigma} \gg u$
overshadows the effect of spatial dispersion completely
(Fig.~\ref{kappaalpha1layer}, right). In this case the result near the lower 
edge of the transmission window is almost the same as in the dispersionless
 model, 
\begin{equation}
T_{\rm max} \approx 4 \kappa (w=1) 
=\frac{4\omega\lambda_{ab}^2k_z}{c\cos\theta}=
\frac{\omega\lambda_{ab}}{\cos\theta\sqrt{a}}~\left(
\frac{a-\tilde{w}}{\tilde{w}-1}\right)^{1/2} , 
\end{equation}
and is mainly determined by  $\sigma_c$.

\section{Crystal with alternating Josephson junctions}

\label{sectionalternate}

For the geometry in Fig.~\ref{geometry}(a) we consider the crystal with two 
 alternating Josephson junctions $l=1,2$
characterized by different critical current densities $J_{0,l}$ and
 two bare plasma frequencies $\omega_{c0,1}$ and 
$\omega_{c0,2}$ related to $J_{0,1}$ and $J_{0,2}$ as described by 
Eq.~(\ref{pl}). We denote $w=\omega^2/\omega_{c0,1}^2$ and 
$\delta=\omega_{c0,1}^2/\omega_{c0,2}^2< 1$. 
In the view of recent experiments  \cite{ourpaper}, we also allow for 
different $c$-axis conductivities $\sigma_{l}$ ($l=1,2$), which are expected 
to vary according to the different tunnel matrix elements in the junctions,
$\sigma_1 / \sigma_2  = \omega_{c0,1}^2 / \omega_{c0,2}^2$, as found for 
La$_{2-x}$Sr$_x$CuO$_4$ \cite{uch}, and which 
are assumed to be frequency independent in the following 
(${\tilde \sigma}_l= 4 \pi \sigma_l / \epsilon_{c0} \omega_{c0,1}$). All other 
parameters of the junctions are assumed to be identical.

\begin{figure}
\begin{center}
\epsfig{file=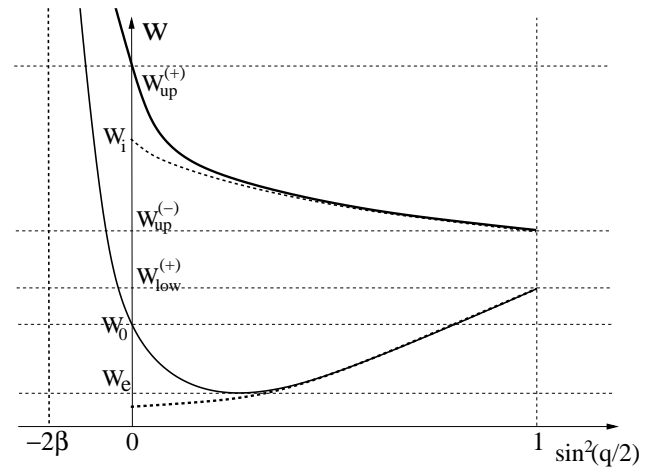,width=0.4\textwidth,angle=-90,clip=}
\end{center}
\caption{
Schematic picture of the dispersion $w(\nu^2) = \omega^2 (\nu) /
 \omega_{c0,1}^2 $ for two alternating junctions ($\sigma_c=0$).  
 The dispersion of the plasma mode at $\theta=0$ (dashed line), i.e. when it 
is decoupled from the electromagnetic wave, is normal in the lower band.  
Its mixing with a decaying electromagnetic wave 
(as shown  by the dashed, vertical line at negative $\nu^2= \sin^2(q/2) 
\approx - 2 \beta $) 
results in two propagating modes (solid) near the lower band edge $w_e$. 
This frequency forms an extremal point with vanishing group velocity as in the 
one band case (cf. Fig.~\ref{schematicmixing1layer} below).
The anomalous dispersion in the uncoupled upper band gives rise to 
one propagating and one decaying mode and a special point $w_i$, where
$q_1 = - i q_2$. The band edges $w_{\rm low,up}^{\pm}$ are defined in the
 appendix A. 
\label{schematicmixing2layer} 
}
\end{figure}

\begin{figure}
\epsfig{file=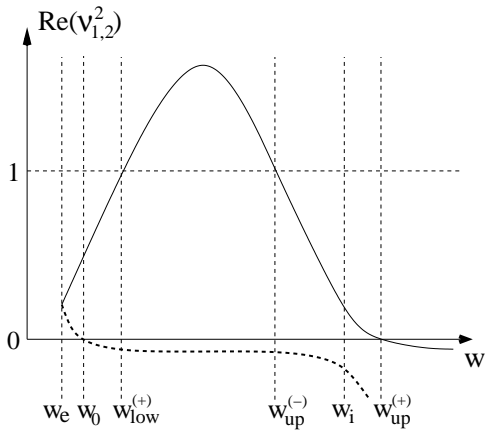,width=0.5\textwidth,clip=}
\caption{Schematic dependence of the squared refraction indices 
 $\nu_{1}^2$ (solid) and
$\nu_2^2$ (dashed) on the squared frequency $w= \omega^2 / \omega_{c0,1}^2$ 
for two alternating 
junctions ($\sigma_l \rightarrow 0$) with the peak positions $w_{e,i}$ 
and the band edges $w^{(\pm)}_{{\rm low,up}}$, as defined in appendix 
\ref{appendixeigen}.  
\label{schematicnu2layer}
}
\end{figure}

The equations inside the crystal are analogous to Eqs.~(\ref{e1})-(\ref{e3})
and the details of their solution are given in appendix \ref{appendixeigen}. 
Here we summarize the main features on the basis of the schematic dispersion 
$w(\nu^2 = \sin^2 (q/2)) $ 
in Fig.~\ref{schematicmixing2layer} and the squared refraction 
indices $\nu_{1,2}^2 (w)$ in Fig.~\ref{schematicnu2layer}.

At perpendicular incidence $a~=~1$ ($\theta~=~0$) the longitudinal plasma mode
is decoupled from the transverse electromagnetic wave, as the incident 
electric field has no component perpendicular to the layers. 
In this case the lower (upper) plasma  bands have a
 normal (anomalous) $c$-axis dispersion (dashed lines in 
Fig.~\ref{schematicmixing2layer}) due to the charge coupling $\alpha$. 

In contrast to this, for $a>1$  the frequency 
$\omega (q)$ increases as
 $q\rightarrow 0$  due to the inductive coupling in both bands (solid lines 
in  Fig. \ref{schematicmixing2layer}). 
For the lower band this can lead  for sufficiently large $\alpha$ 
to an extremal point $w_e$
at the lower band edge  as in the case of identical layers, where
(for $\sigma_l =0$) two modes with  real $q$ exist,
 while  near the upper band edge $w_{\rm low}^{(+)}$ 
one mode propagates and the other decays. 
In the upper band there is one real and 
one imaginary  solution everywhere in the band due to the anomalous 
dispersion, and 
according to Eq.~(\ref{tmaxi}) in the general section II 
the maximal transmission is at $w_i$, where $q_1^2=-q_2^2$. 

All special frequencies mentioned in Figs. \ref{schematicmixing2layer}
and \ref{schematicnu2layer} are explicitly expressed by microscopic 
parameters in appendix \ref{appendixeigen}. For the frequencies $w_{e,i}$  
of the resonance maxima we obtain approximately
\begin{eqnarray}
w_{e,i}&\approx& (1+\delta)(1+2\alpha)/2\delta \mp \label{wmax2layer} \\
&&[(1+\delta)^2(1+2\alpha)^2-4\delta(1+4\alpha )]^{1/2}/2\delta. \nonumber
\end{eqnarray}

As for identical layers the optical properties are dominated by the mode with 
smaller $|n_p| = c |\nu_p | / s \omega$ and significant deviations from the
one mode Fresnel regime occurs at $|n_1|\approx |n_2|$.

\begin{figure}
\epsfig{file=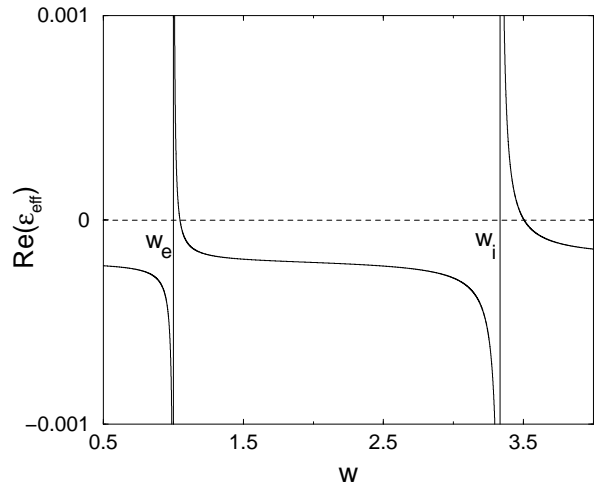,width=0.37\textwidth,clip=,angle=-90}
\caption{The real part of the effective dielectric function
 $\epsilon_{\rm eff} = \kappa^2$ without explicit dispersion ($\alpha=0$, 
$\delta = \omega_{c0,1}^2 / \omega_{c0,2}^2 = 
 0.3, a=1.1, \beta = 10^{-4}, {\tilde \sigma}_l=0$). In this limit
the dielectric function is directly related to the refraction index and the
wavevector of the single excited mode, $\epsilon_{\rm eff} \propto n_0^2
\propto k_z^2$, see Eqs.~(\ref{kappaonemode}) and (\ref{refractionpure}).  
The poles at the lower band edges, where the averaged ${\tilde \epsilon}_c 
(\omega) =0$  vanishes, indicate the appearance of the special points 
$w_e$ and $w_i$ in the two mode theory, cf. Fig.~\ref{schematicnu2layer}
for the corresponding case 
$\alpha \neq 0$ and Fig.~\ref{poleschematic} for the general picture.
\label{epseff}
}
\end{figure}

Keeping only the solutions with smallest $|\nu_p|$ near $w_i$ and $w_e$, e.g. 
$\nu_2$ ($\nu_1$) for $w<w_i$  ($w>w_i$) in the upper band, we obtain in the
limit $\alpha \rightarrow 0$ a pole in $\nu^2 \sim q^2 \sim \epsilon_{\rm
eff}$, as can be seen from Fig.~\ref{epseff}. 
This is an explicit microscopic confirmation of the general expectation 
that critical frequencies, where $|\nu_1|=|\nu_2|$, appear, if there is a pole 
in $k_z^2 \sim \epsilon_{\rm eff}$. For oblique incidence the 
singularities in $\epsilon_{\rm eff} (\omega)$ coincide with the zeros of 
the averaged ${\tilde \epsilon} (\omega)$ introduced in 
Eq.~(\ref{dielecaver}) with 
$\epsilon_{cl} = \epsilon_{c0} (1 - \omega_{c0,l}^2 / \omega^2 )$ . 
This can be expected from a macroscopic treatment, where the spatially 
averaged ${\tilde \epsilon}$ is introduced in Eq.~(\ref{refractionpure}). 

The regularization of the poles in $\epsilon_{\rm eff}$ is seen by comparing 
the behaviour of ${\rm Re}(\nu_p^2) \sim {\rm Re} (k_{zp}^2)$ for 
$\alpha \neq 0$ (Fig.~\ref{schematicnu2layer}) and of 
${\rm Re} (\epsilon_{\rm eff}) \sim k_z^2$ for $\alpha =0$ (Fig.~\ref{epseff}) 
near $w_e$  and $w_i$ with the schematic picture in Fig.~\ref{poleschematic}. 

In sect. V it will be shown that a situation, where a second mode contributes
in a similar way as near the point $w_i$ in the upper band, can also develop 
from a pole in the dispersionless dielectric function without explicit spatial
 dispersion, e.g. for $\alpha =0$, due to the 
intrinsic atomic structure within the unit cell, see Fig.~\ref{wparallel}. 

Also like in the single junction case, the transmission into the crystal in 
the lower band is only significant, if both excited modes are propagating 
into the crystal.  
Consequently, the width of the 
resonance $\Delta_{\rm low}\approx w_0-w_e$ ($\Delta_{\rm up} \approx 
w_{\rm max}^{+} - w_i $) in transmission $T(w)$ in the lower (upper) band 
are considerably smaller than the band width of the allowed eigenmodes in the
crystal, $w_{\rm low}^{+}-w_e$ or $w_{\rm up}^{+}-w_{\rm up}^{-}$
respectively, see Fig. \ref{schematicnu2layer}. 

As derived in Appendix \ref{appendixeigen} the additional boundary
condition near the special points $w_e$ and $w_i$ is analogous to the case 
of identical junctions Eq.~(\ref{MBC}) and reflects the 
fact that on the surface one neighboring junction is missing. 
In leading order of $\beta^{3/4} \sim \epsilon_a/n_p^2$ and 
$b/\beta^{1/4} \sim 1/n^2$ we obtain
\begin{equation}
P_{m=-1,2} =  \sum_{p=1,2} \gamma_{p} P_2(q_p) \exp(-iq_p)=0.
\label{111}
\end{equation} 
Thereby $P_{m=-1,2}=\int_{-s}^{0} P_{z2}  dz$ is the average of the $l=2$
component $P_{z2}$ of the macroscopic polarization vector in the missing 
junction in the cell $m=-1$, 
$(P_1 (q), P_2 (q))$ denotes the eigenvector of the excited mode and
$\gamma_p$ describes the relative amplitude of the excited modes $p=1,2$, 
see Eq.~(\ref{Avect}).
This microscopic result gives an a posteriori justification 
of the phenomenological ABC in Eq.~(\ref{ABC}) for the multimode case, 
where the length scale $\ell = -2s$ is identified with the lattice constant
in $c$-direction. This shows in particular that the macroscopic approach
is possible, if and only if multicomponent local polarizations 
$\tilde{P}_{1z},\tilde{P}_{2z}$ inside the unit cell are
introduced. Expanding ${\tilde P}_2 (q):= P_2 (q)\exp(-iq) \approx P_2 (0) 
( 1 - i q / 2) $ with the help of Eq.~(\ref{Avect}) for 
$P_2 (q) \approx P_2 (0) ( 1 + i q )$
and taking into account the doubled unit cell in $q=2sk_z$, we obtain  
Eq.~(\ref{ga}) with
the effective parameter $\xi  = - \omega s / c $. 
It is pointed out that the same result for the amplitude ratio of the excited 
modes as in the single layer case is reached here in a nontrivial way 
by an interplay of the lattice constant $\ell=2s$ and the internal 
structure of 
the eigenmodes contained in $P_l(q)$.  

\begin{figure}
\epsfig{file=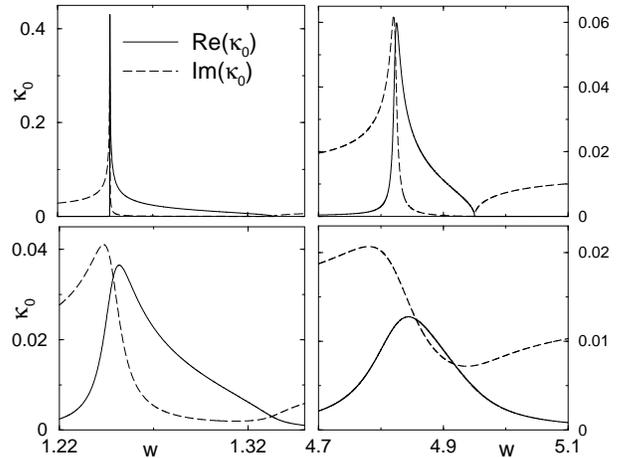,width=0.35\textwidth,clip=,angle=-90}
\caption{ 
Real and imaginary part of $\kappa_0= \kappa 
(\cos\theta \epsilon_a \omega s / c  ) $  for alternating 
junctions ($\alpha=0.2$, $a=1.1$, 
$\delta= \omega_{c0,1}^2 / \omega_{c0,2}^2 = 0.3$, $\beta=10^{-4}$) near
$w_{e} \approx 1.244$ (left) and $w_{i} \approx 4.823$ (right)  
for different quasiparticle dissipation 
${\tilde \sigma}_1 = \delta {\tilde \sigma}_2 = 10^{-6}$ (above) or
$0.005$ (below). \label{kappa2layer} }
\end{figure}

Now we are in the position to calculate the reflection and transmission 
coefficients $R$ and $T$ near the resonances in leading order in $\beta^{3/4}
\sim \epsilon_a / n_p^2$ and $b/\beta^{1/4} \sim 1/n^2$, where  $\kappa$ is  
given by  
\begin{eqnarray}
\kappa &=&
\frac{1}{\epsilon_a  \cos\theta}~
\frac{n_1n_2[{\tilde P}_2(q_1)n_2-{\tilde P}_2(q_2)n_1]}
{{\tilde P}_2(q_1)n_2^2-{\tilde P}_2(q_2)n_1^2} 
\label{kap2}
\end{eqnarray}
and shown in Fig.~\ref{kappa2layer}. Due to the lattice constant
 $2s$  the refraction indices of the bulk eigenmodes
are here $n_p = c k_{z,p} / \omega = c \nu_p / s \omega $.  
This result reduces to Eq.~\ref{gen}, when expanding ${\tilde P}_2 (q)$, and
the results of the general Sect.~\ref{macrosection} can be used.

\begin{figure}
\epsfig{file=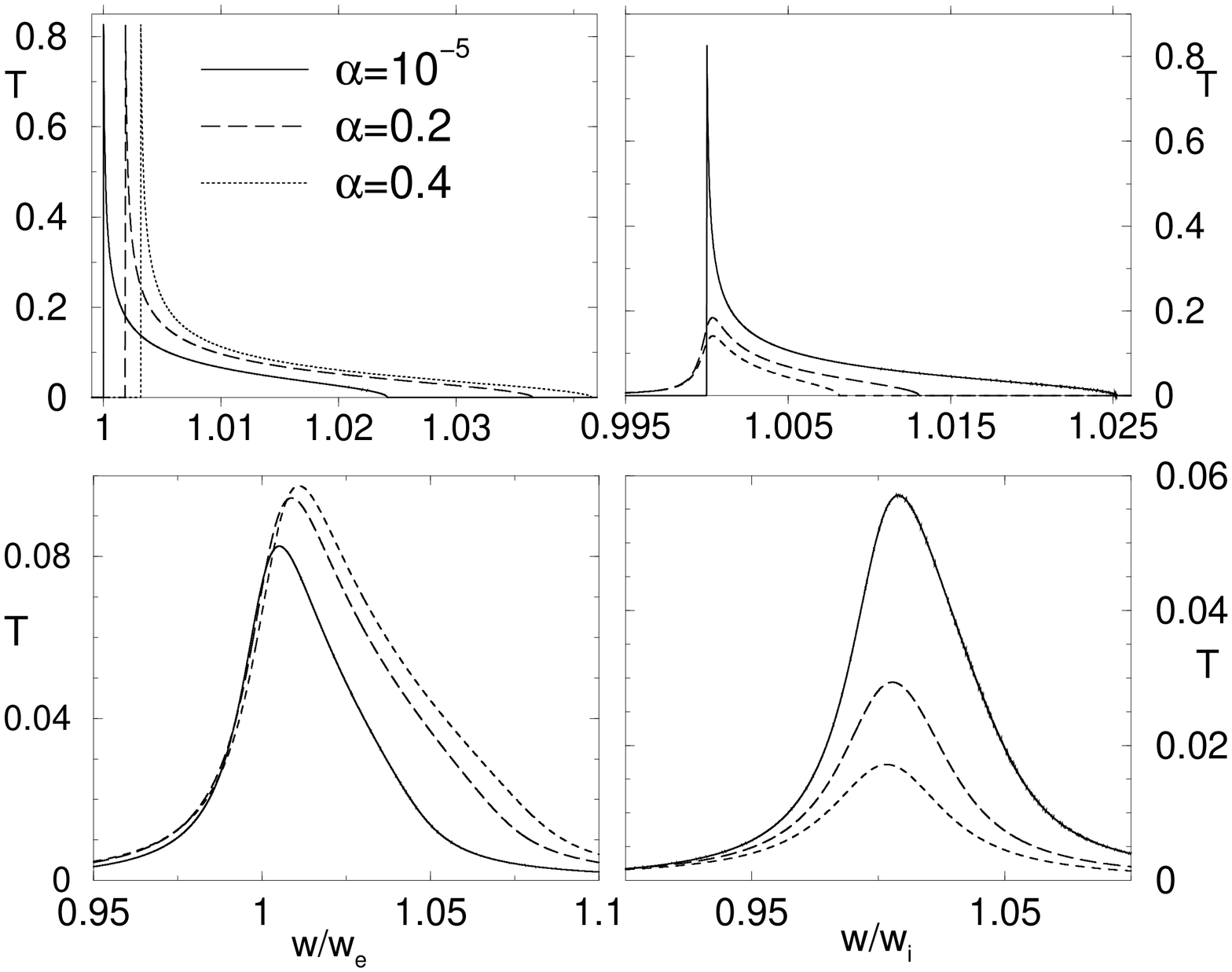,width=0.38\textwidth,clip=,angle=-90}
\caption{
Transmissivity $T$ near the lower (left) and upper (right) plasma bands 
with the 
frequency axis normalized to $w_{\rm low}$ ($w_{\rm up}$) respectively.
Parameters: 
$\delta =0.3$, $\beta=10^{-4}$, $ \epsilon_a \xi \cos \theta 
= 1$, different conductivities in the plots 
above (${\tilde \sigma}_1 = \delta {\tilde \sigma}_2 = 10^{-8}$, $a=1.05$) 
and  below (${\tilde \sigma}_1 = \delta {\tilde \sigma}_2 =0.01$, $a=1.1$) 
and varying $\alpha$ (see plot). 
\label{T2layerplot} 
}
\end{figure}


The lower band is  similar to the case of identical junctions in the sense 
that in the transmission window $w_e \le w \le w_0$ two propagating 
modes are excited and we obtain the same maximal transmission coefficient 
$T_{\rm max,low}=T_{\rm e,max}$ (cf. Eq.~(\ref{Tmax1macro})).  

In the upper band   we get from the general Eq.~(\ref{tmaxi})
for small dissipation ${\tilde \sigma}_l \ll u_{\rm up}$ at $w=w_i$ 
\begin{equation}
T_{{\rm max,up}} = T(\omega_i)   =  \frac{2 \lambda_{ab}^{3/2}\epsilon_{a0}}
 {\lambda_c(s\epsilon_{c0})^{1/2}\cos\theta }
 \left[\frac{(a-1)L_{\rm up}}{8\alpha^2 a}\right]^{1/4},
\end{equation} 
which is smaller by  the factor $(s/\lambda_{ab})^{1/2}$ than 
$T_{{\rm max,low}}$ ($L_{\rm up}=  w_i (1+\delta) - 2 - 8 \alpha$).
 This can be seen in the 
upper part of Fig.~\ref{T2layerplot}, where for low ${\tilde \sigma}_l \ll 
u_{\rm low,up} \ll 1$ (see definitions in App. \ref{appendixeigen})
 the upper plasma resonance is considerably suppressed by 
increasing $\alpha$, while the lower band is weakly affected.  
This suppression can be understood physically by the fact that at the surface 
the energy of the incident wave is distributed between a propagating wave and 
a decaying (and finally reflected) 
one and is therefore less efficiently transmitted in the crystal 
than in the lower band, where the two excited modes are propagating.  
Physically, the eigenvectors near $q\approx 0$ in the lower 
(upper) band involve in phase 
(out of phase) plasma oscillations and consequently external
long wave length radiation couples more efficiently to the excitations in the
lower than those in the upper band. 

The difference between the values of $T_{{\rm max,low}}$ and $T_{{\rm max,up}}$
decreases as dissipation increases, see Fig.~\ref{T2layerplot} below. 
It vanishes in the Fresnel limit, for which  
$(4\pi\sigma_1/\omega_{c0,1} \epsilon_{c0})(\lambda_{ab}/\alpha s)$ 
becomes much larger than unity. 

In oblique incidence the suppression of the peak in the 
upper band is quite limited to 
systems with very low dissipation and perfect crystal structure 
and might be difficult to observe in 
SmLa$_{1-x}$Sr$_{x}$CuO$_{4-\delta}$. 
Instead of this, a quite high ratio of the peak amplitudes has been observed
in this material for incidence parallel to the layers
\cite{marel,marel2,shibata,kakeshita,pim}, 
see below and Ref.~\cite{marel4,ourpaper}.

\section{Incidence of light parallel to the layers }

\label{sectionparallel}

In this Section we discuss the reflectivity for incidence parallel to 
the layers, cf. Fig.~\ref{geometry}(b) for $\theta=0$, 
in the crystal with two alternating junctions, when the explicit spatial
dispersion, i.e. the dependence on the wavevector $q$, is negligible. 

We will confirm microscopically  the
breakdown of the macroscopic Fresnel approach using the effective
dielectric function ${\tilde \epsilon}_c$, Eq.~(\ref{dielecaver}), when the 
wavevector $|k_{xp}|$ of the excited modes becomes large and the group
velocity is small,  cf. Eq.~(\ref{vgroup}).
This happens near the pole $\omega_{\rm pole}$ of $\tilde{\epsilon}_c$, 
which coincides with the upper edge of the lower plasma resonance in the
reflectivity (cf.~Fig \ref{Rparallel}). This frequency is sometimes associated 
with the excitation of a so called "transverse" mode \cite{marel3,marel4}, although
all the modes excited in the plasma bands are transverse in this geometry. 
For simplicity we will present here the formulas for $\alpha=0$, 
the general results are given in Appendix \ref{appendixparallel}. 

Physically, the conventional theory is insufficient, because it 
averages the Eqs.~(\ref{first})~-~(\ref{last}) within the unit cell and
neglects the electric field components parallel to the layers, in order to
arrive at the response function ${\tilde \epsilon}_c$ for the averaged field 
$E_{z,m,{\rm av}} = \int_{ms}^{(m+2)s} E_z dz$. This corresponds to neglecting
the average 
$\int_{ms}^{(m+2)s}dz \partial_z B_y = B_y [(m+2)s] - B_y (ms)$ and the average
of $\partial_z E_x$ respectively, i.e. to setting 
$k_z=0$ in the Eqs.~(\ref{max1})~-~(\ref{max3}).
This assumption is justified away from $\omega_{\rm pole}$, where the 
wavevector $|k_x|$ is small, as the gradient of the electric field vanishes,
if the charge density on the layers is slowly varying, $ \partial_z E_x \sim
|k_x|$. On the other hand, at $\omega_{\rm pole}$ the
charge density varies on atomic scales, the intra-junction mode with
polarization of the electric field in $x$-direction is excited strongly and 
the basic
assumption of the averaged theory is invalid.

\begin{figure}
\begin{center} 
\epsfig{file=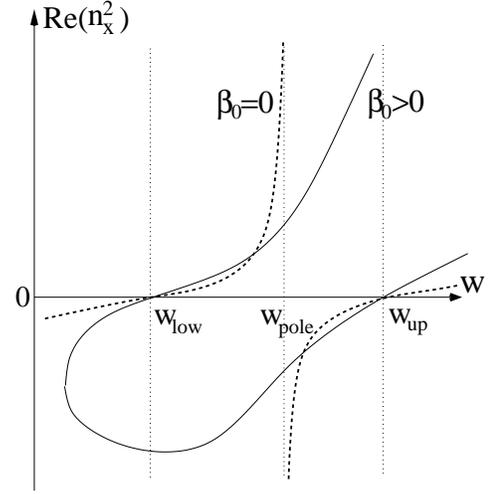,width=0.35\textwidth,clip=,angle=0}
\end{center}
\caption{
Schematic refraction index $n_x^2 (w)$ without dispersion $\alpha=0$, but 
with (solid line $\beta_0 = s^2 / 2 \lambda_{ab}^2 >0$) 
or without (dashed, $\beta_0=0$) accounting for
the intrinsic inhomogeneity in the unit cell. The latter implicit 
spatial dispersion corresponds to 
 the excitation of the mode with electric field polarization parallel to
the layers.
 The frequencies $w (k_x=0) = w_{\rm low, up}$ form the 
plasma edges in the reflectivity $R_{\parallel}$. 
We can also interpret the plot for $\beta_0 =0$ as the averaged dielectric
function ${\tilde \epsilon}_c \sim k_x^2 \sim n_x^2$ (cf. Eq.~ (\ref{r})) as a
function of $w$. Then the pole in 
${\tilde \epsilon}_c (w)$ at $w_{\rm pole}$ in the one mode approach
 indicates the appearance of a special frequency for $\beta_0 \neq 0$, 
where $n_{x1}^2 = - n_{x2}^2$, which is similar to the general picture in
Fig.~\ref{poleschematic} and  the upper band in oblique incidence, cf. 
Fig.~\ref{schematicnu2layer}. 
\label{wparallel}
}
\end{figure}

\begin{figure}
\begin{center} 
\epsfig{file=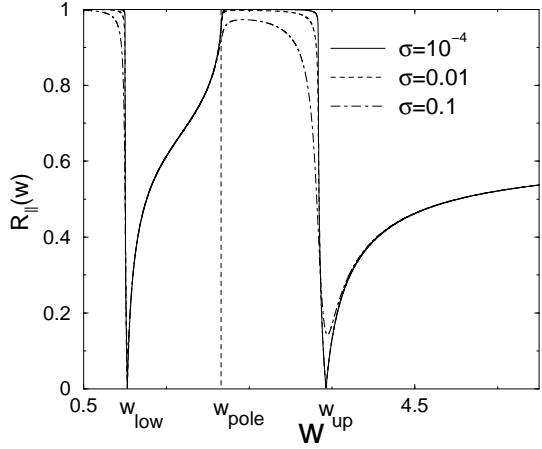,width=0.35\textwidth,clip=,angle=-90}
\end{center}
\caption{Reflectivity $R_{\parallel}
=|1-\kappa_{\parallel}|^2/|1+\kappa_{\parallel}|^2$ in
 parallel incidence  for  $\alpha=0$, $\delta= \omega_{c0,1}^2 /
 \omega_{c0,2}^2 = 0.3$,  $\epsilon_{c0}=19$, 
 $\beta_0= s^2/2 \lambda_{ab}^2 = 10^{-4}$ 
as in the cuprates and different conductivities 
${\tilde \sigma}_1= \delta {\tilde \sigma}_2$ (see plot). 
\label{Rparallel}
}
\end{figure}

A more careful averaging of the Eqs.~(\ref{first})-(\ref{last}) within the
junctions rather than the unit cell leads to relation between average 
electric fields inside junctions ($\alpha=0$, $\sigma_l =0$, for the
general case see App. \ref{appendixeigen}) 
\begin{equation}
\left[ \left( \frac{c k_x}{\omega}  \right)^2 
\frac{1}{2+\beta_0} \left( 
\begin{array}{cc} 1+\beta_0 & 1 \\ 1 & 1+ \beta_0  \end{array}  \right)
- \underline{\underline{\epsilon}}  
\right]
\left( \begin{array}{c} E_{z,1} \\ E_{z,2} \end{array}  \right)
= 0 . 
\label{2layerdispersion}
\end{equation}
Here the dielectric tensor $\underline{\underline{\epsilon}}$ is given 
as $\epsilon_{ll} = \epsilon_{cl} =
\epsilon_{c0} (1 - \omega_{c0,l}^2 / \omega^2)$,
$\epsilon_{12}=\epsilon_{21}=0$ and $\beta_0=a \beta = s^2 /(2
\lambda_{ab}^2) \ll 1$ accounts for the  coupling of the averaged electric
fields $E_{z,1,2}$ in the junctions of type $l=1,2$ via the
electric field component $E_x$. The latter is weak, $\sim \beta_0$, due to 
the strong anisotropy of the material. 
For $\beta_0 =0$ one eigenmode of Eq.~(\ref{2layerdispersion})
corresponds to the solution in the averaged theory determined by 
${\tilde \epsilon}_c$, $ c^2 k_{x1}^2 = \omega^2
{\tilde \epsilon}_{c}$, and its eigenvector obeys $D_z = \epsilon_{c1} E_{z1} = 
\epsilon_{c2} E_{z2}$ as it is assumed in macroscopic electrodynamics. Consequently, 
near the lower band edges $\omega_{c0,l}$ only the plasmon in the junction of 
type $l$ is excited. The other mode has an eigenvalue $1/k_{x2}^2 =0$ and 
corresponds to an out of phase mode with the eigenvector $E_{z1} = - E_{z2}$,
which is not excited by a homogenous incident beam. 
Accounting for the excitation of the electric field components parallel
to the layers, at $\beta_0 > 0$,  both modes mix and the singularity at 
$\omega_{\rm pole}$ is removed. This is shown schematically in 
Fig.~\ref{wparallel} 
and is a consequence of the presence of two 
junctions in the unit cell (implicit spatial dispersion), even in
the absence of $c$-axis coupling ($\alpha =0$ or $q=0$).


Including the dissipation due to $\sigma_l$, in 
 the case $q \ll b$ ($\Leftrightarrow {\rm sin} \theta \ll 1$)  of
 parallel incidence the dispersion is given by  Eq.~(\ref{dispapp2}), 
 and the general solutions 
$n_{xp}= c k_{xp} / \omega$ are presented in Eqs.~(\ref{kxgen}) in the 
appendix. Away from the pole $w_{\rm pole}$ these solutions
can be expanded in leading order in $\beta_0$ and we obtain the usual wave
 with the refraction index $n_{x1} = c  k_{x1} / \omega$
 corresponding to the average $\tilde{\epsilon}_c(\omega)$
(for $\alpha=0$):
\begin{eqnarray}
&&1-\frac{1}{a_1}=\frac{c^2k_{x1}^2}{\omega^2\epsilon_{c0}}=
\frac{\tilde{\epsilon}_c(w)}{\epsilon_{c0}}=
\frac{\delta (w-w_{\rm low}) (w -w_{\rm up})+ i S}{w \delta (w-w_{\rm pole})+i
 S_1}, 
\nonumber \\   
&&S_1= (1/2) w^{3/2} \delta  ({\tilde \sigma}_1 + 
{\tilde \sigma}_2) ,  \label{r}   \\
&&S=w^{1/2} [ \tilde{\sigma}_1 (\delta w-1) + \delta \tilde{\sigma}_2 
(w-1)], \nonumber
\end{eqnarray}
The zeros of $k_{x1}$ are at the plasma edges $w_{\rm low, up}$, 
\begin{eqnarray}
w_{\rm low,up}&& =(1+\delta)(1+2\alpha)/2\delta \mp \label{100} \\
&&[(1+\delta)^2(1+2\alpha)^2-4\delta(1+4\alpha (1-\beta z_0 ) )]^{1/2}/2\delta. \nonumber
\end{eqnarray}
For $\beta_0=0$ this corresponds to the single excited mode $k_x^2 \sim 
{\tilde \epsilon}_c$ and we see
 from Fig.~\ref{wparallel} (dashed line) 
that $|n_{x1}|$ becomes large at the pole $w_{\rm pole}=(1/2+2\alpha)
(1+\delta^{-1})$.  
The discrete layered structure 
($\beta_0 \neq 0$) results in the regularization of the pole and its
transformation into a special frequency $\omega_i$, 
where $n_{x1}^2 = - n_{x2}^2$ without dissipation, 
see Fig.~\ref{wparallel} (solid). 
This is similar to the behavior in the upper 
plasma band in oblique incidence, see Fig.~\ref{schematicmixing2layer}, where 
a pole in the one mode Fresnel dielectric function $\epsilon_{\rm eff}$ is 
transformed into the special point $w_i$. 
There it was a consequence of explicit spatial dispersion ($\alpha \neq0$), 
while now the second solution $n_{x2}$ appears  due to the atomic 
structure within the unit cell even at $\alpha = 0$.

Away from the pole $w_{\rm pole}$ we get (for $\alpha=0$)
\begin{equation}
1-\frac{1}{a_2}= \frac{c^2k_{x2}^2}{\omega^2\epsilon_{c0}}
=\frac{ w \delta ( w - w_{\rm pole} ) + i S_1}{ (\beta_0/2) w^2 \delta},
\end{equation}
and $|k_{x2}| \sim O(1/\beta_0)$ is large in comparison with
$|k_{x1}|$. As the solution with 
the smallest refraction index determines the optical properties, the wave 
with wave vector $k_{x2}$ can therefore be neglected everywhere except at
$w_{\rm pole}$, 
where $k_{x2}$ is small at weak dissipation and 
the general Eqs.~(\ref{kxgen}) have to be used. 

Let us now find the solutions for the magnetic and electric fields 
inside the crystal which determine the reflection coefficient 
$R_{\parallel} = |(1-\kappa_{\parallel})/(1+\kappa_{\parallel})|^2$. 

The solution for $B_y$ at $x<0$ consists of the
incident and  reflected waves $B_{y}^{\rm in}$ and $B_{y}^{\rm ref}$,
 which are homogeneous in the $z$-direction, 
and the wave ${\cal B}_y$ with $k_z \neq 0$, which is excited due to the 
inhomogeneity of the crystal in $z$-direction and which is localized near the
surface: 
\begin{eqnarray}
B_y(x)&=&B_{y}^{\rm in}\exp(ik_xx)+B_{y}^{\rm ref}\exp(-ik_xx) \nonumber \\
&+&\sum_{k_z\neq 0}{\cal B}_y (k_z)
\exp(ik_zz-i\tilde{k_x}x), 
\label{by}
\end{eqnarray}
where $c^2(\tilde{k}_x^2+k_z^2)=\omega^2$. The solution at $x>0$ is given 
by Eq.~(\ref{def1}), when introducing $k_{xp}$ explicitly by substituting 
$\sin \theta \rightarrow - c k_x / \omega$ and taking into account the 
superposition of the two solutions $p=1,2$.  

In addition to this, we need an additional boundary condition, in order to 
determine the ratio, in which the modes $p=1,2$ are excited. 
At $\theta=0$ the in-plane currents $J_{x,m} = 
i \omega_a^2 E_x (z=ms) / [4 \pi \omega]$ and  consequently the $E_x$ 
components inside the layers $m$ vanish at $x\rightarrow 0$.
This is equivalent to Pekar's boundary condition 
$P_x=0$, which turns out to be sufficient in this case due to the absence of 
extremal points, where $n_1 + n_2 \approx 0$ (cf. Fig.~\ref{wparallel}). 

As worked out in appendix \ref{appendixparallel} this leads to the reflection
 coefficient $R_{\parallel}$, Eq.~(\ref{kappagen}), where ($\alpha = 0$)
\begin{eqnarray}
\kappa_{\parallel}&=& \sqrt{\epsilon_{\parallel}^{\rm eff}} = 
\frac{a_1+a_2Z}{a_1n_{x1}+a_2n_{x2}Z},
\label{kappaparallel0}
 \\
Z&=&-\frac{(1-f_1)(1+f_2)(a_1-1)k_{x2} a_2 }
{(1-f_2)(1+f_1)(a_2-1)k_{x1} a_1}, \\
f_p&=& \frac{({\tilde w}_1-1) (2 a_p + \beta_0)  -
(a_p-1)(a_p+\beta_0)}{(a_p-1)(a_p + \beta_0 )}.
\end{eqnarray}

\begin{figure}
\begin{center} 
\epsfig{file=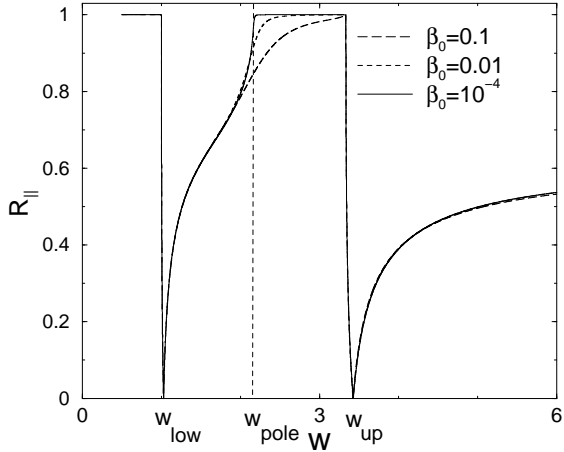,width=0.35\textwidth,clip=,angle=-90}
\end{center}
\caption{Reflectivity $R_{\parallel}
=|1-\kappa_{\parallel}|^2/|1+\kappa_{\parallel}|^2$ in parallel incidence  
for  $\alpha=0$, $\delta=0.3$, ${\tilde \sigma} = 10^{-6}$,  
$\epsilon_{c0}=19$ and different $\beta_0 = s^2 / 2 \lambda_{ab}^2$ 
(see plot). It is seen that, when the second solution is taken into 
account,  at $\beta_0 \gg 0.0001$, 
the reflectivity drops near the pole 
$w_{\rm pole}$ of the averaged dielectric function 
${\tilde \epsilon}_c$, where $n_{x1}^2 = - n_{x2}^2$.
\label{Rparallelbeta}
}
\end{figure}

In Fig.~\ref{Rparallel} the reflectivity $R_{\parallel}(w)$  
is shown for different conductivities ${\tilde \sigma}_1 = \delta 
{\tilde \sigma}_2$ and a value $\beta_0=10^{-4}$ appropriate for high 
temperature superconductors.
The resonances in the lower and upper plasma bands are asymmetric and have
a sharp lower edge at $w_{\rm low,up}$. The upper edge of the lower band 
is given by the pole $w_{\rm pole}$, where in the conventional one mode theory 
${\tilde \epsilon}_c$ becomes negative and the single excited mode with an
imaginary wave vector decays (cf. Fig.~\ref{wparallel}). 

In Fig.~\ref{Rparallelbeta} the effect of the discrete layered structure 
($\beta_0\neq 0$) on the reflectivity $R_{\parallel}$ in parallel 
incidence is shown. 
At the special point $w_{\rm pole}$, where  the second solution
$n_{x2}^2 = - n_{x1}^2$ becomes relevant,  the reflectivity 
$R_{\parallel}$ drops for large $\beta_0$ and the lineshape is modified.
This behavior is 
similar to the resonance at $w_i$ in the upper plasma band for oblique 
incidence due to spatial dispersion, see Sect. IV. 
This modification of the JPR lineshape  is beyond the conventional one
mode Fresnel approach, which is valid away from  $ w_{\rm pole}$, 
in particular near the plasma resonances.

Therefore for the interpretation of the main peak amplitudes the simplified 
effective dielectric function ${\tilde \epsilon}_c$ is sufficient and has 
been used in \cite{ourpaper} to extract the parameter $\alpha \approx 0.4$
from the experimental loss function in SmLa$_{1-x}$Sr$_{x}$CuO$_{4-\delta}$.
In contrast to \cite{marel4} the dissipation was introduced here
microscopically in the quasiparticle currents and it is taken into account
that the quasiparticle conductivities alternate,
 $\sigma_1 / \sigma_2 = \omega_{c0,1}^2 / \omega_{c0,2}^2$, 
in the same way as the critical current densities 
and plasma frequencies $\omega_{c0,l}$.  Correctly accounting for dissipation 
is crucial for a quantitative 
interpretation of the experimental loss function. 
As the parameter $\alpha$ can be extracted independently from the magnetic 
field dependence of the plasma resonances, see \cite{ourpaper}, 
this is also a way to determine the $c$-axis conductivities $\sigma_l$. 
Both ways to extract $\alpha \approx 0.4$ from far-infrared data are well 
compatible with the ARPES measurements \cite{norman}. 

\section{Conclusions} 

In conclusion, the effect of spatial dispersion and the atomic structure 
 on the optical properties 
of stronly anisotropic 
uniaxial crystals has been studied in general, taking as a generic example 
the Josephson Plasma Resonance in stacks of identical or alternating 
junctions. 

Thereby, multiple eigenmodes, propagating or decaying, 
are excited  by incident light, which interfere with each other. This 
intrinsic birefringence can be detected in transmission by oscillations
with respect to the sample thickness or the splitting of the incoming (laser) 
beam (cf. Sect. II.C). 

In contrast to the usual assumption that the effect of dispersion or of the
atomic structure on  
optical characteristics is strongly suppressed $\sim s/\lambda \ll 1$,
as the wavelength $\lambda$ of light is much larger than the lattice 
constant $s$, we showed that near resonance frequencies the reflectivity may
differ significantly from the conventional Fresnel formulas, if dissipation
and disorder are weak.

Near extremal frequencies $\omega_e$, where the group velocity $v_g =
\lambda_g / \omega $ vanishes, the stopping of the wave packet makes the 
propagating light sensitive to short length scales $\lambda_g$. 
As a consequence, for oblique incidence the transmissivity into the crystal 
cannot be expressed by the bulk dielectric function alone and the 
amplitude of the resonance near $\omega_e$ crucially depends on the atomic 
structure of the crystal. This additional damping due to the $c$-axis coupling
$\alpha$ for low dissipation is shown in Fig. \ref{kappaalpha1layer}.
In contrast to this, the width of the 
resonance in transmission is not affected by the $c$-axis charge coupling,
 but is rather determined by the angle of incidence. 
 
These extremal points $\omega_e$ may appear, whenever an optically active 
crystal mode 
with normal (anomalous) dispersion is mixed with a propagating (decaying) 
electromagnetic wave. For these results it was crucially to realize that the
 resulting 
two eigenmodes with normal and anomalous dispersion have wave vectors and 
refraction indices with opposite sign near $\omega_e$ 
in order to preserve causality. 

In addition to this, for a crystal with several optical bands we predict 
different amplitudes of the resonance transmission into 
bands, which are characterized by different types of dispersion and which are
 equivalent in a dispersionless theory. 
When inside the crystal one mode is propagating and the other one is decaying, 
the maximum of $T$ is at frequencies $\omega_i$, where the relation $n_1 = - i
n_2$ for the refraction indices holds. At these frequencies 
 the peak amplitude of $T$ is strongly suppressed in comparison 
with bands, where the two excited modes are propagating 
(Fig. \ref{T2layerplot}), provided that the dissipation is low. 
This provides the unique opportunity to extract 
microscopic information about the eigenvectors of the excited modes from the 
line shape in optical experiments.

For incidence parallel to alternating layers a second mode is excited even 
without explicit spatial dispersion (${\bf k}$-dependence of $\epsilon_c$)
 due to the intrinsic inhomogeneity within the
unit cell. Near the pole of the effective 
dielectric function at the upper edge of the lower plasma band a special point 
appears, where $n_{x1}^2 = - n_{x2}^2$. For an 
appropriate choice of parameters this can modify the lineshape of the 
resonance. 

This behavior near $\omega_e$ and $\omega_i$  cannot be obtained in the one
mode approach without dispersion. The only intrinsic indication for the
breakdown of the conventional Fresnel theory is 
the appearance of poles in the effective dielectric function 
$\epsilon_{\rm eff}$, see the schemcatic Fig.~\ref{poleschematic}. 
There the excited wave vectors 
$k^2 \sim \epsilon_{\rm eff}$ are large, the group velocity is small, 
cf. Eq.~(\ref{vgroup}), and concomitantly  small atomic length 
scales become important (cf.~Fig.~\ref{epseff} and \ref{wparallel} 
for oblique and parallel incidence). 

These features were demonstrated explicitly for the JPR with identical and 
different alternating junctions, but they are general for any modes,
e.g. for optical phonons with anomalous dispersion in insulators, which form a 
polariton branch with an extremal point, see Fig.~\ref{polariton}. 
However, the condition of weak dissipation and a perfect crystal structure 
are crucial to observe deviations from the Fresnel regime.

For the JPR this theory was used to extract the parameter 
$\alpha \approx 0.4$ from the optical data obtained for
SmLa$_{1-x}$Sr$_{x}$CuO$_{4-\delta}$ with two different alternating 
intrinsic Josephson junctions between the CuO$_2$ single layers \cite{ourpaper}. 
This value corresponds to an electronic
compressibility, which is unrenormalized by the
 interaction, while for multilayer cuprates a smaller value of $\alpha$ 
is expected. This result is compatible with the ARPES 
measurements \cite{norman} 
and gives an important input parameter for the coupled 
Josephson dynamics in the stack. Thereby the correct treatment of the  
$c$-axis conductivities in different junctions is essential for a quantitative 
interpretation. 

It is also pointed out that spatial dispersion provides a way to stop light 
in a crystal, which is dual to previous proposals based on the frequency 
dispersion of the medium, see Sect. II.D. 
From the application point of view, this suggest future magneto-optical 
devices (using e.g.~the JPR) for storing light coherently, as it is 
required in an optical quantum computer. By imprinting a group velocity
profile with the help of an inhomogeneous external magnetic field,  event
horizons with respect to the propagation of light can be created in a solid.

To summarize, possible experiments to demonstrate the effect of spatial 
dispersion on the optical properties of solids include 
the demonstration of: (a) intrinsic 
birefringence and beam splitting,  (b) stopping (delaying) light pulses, 
(c) the relative amplitude of bands with a different number of propagating 
excited modes and (d) the intrinsic damping of peak amplitudes in materials
with negligible dissipation and disorder. 

From a general point of view, these results shed new light on the 
long standing question of the treatment of spatial dispersion for optical 
properties of solids and provide the first microscopic derivation of the ABC
as suggested in \cite{agr}. It is expected that the 
phenomenological results presented here can have wide implications for the
 interpretation of resonance amplitudes and lineshapes in optical experiments, 
 especially near frequencies $\omega_e$ or $\omega_i$, which appear near poles 
of the conventional dielectric function.  
Moreover, the method to obtain the  parameter $\ell$ microscopically by
considering the difference between the hypothetical bulk and the real equation
of motion for surface degrees of freedom, can be generalized to other systems. 
In particular for optical phonons (polaritons) in insulators \cite{ivopaper}
and photonic crystals \cite{photonic,photonicdisp1}
 some of the above deviations from the conventional
Fresnel theory can be expected. 

The authors thank G. Blatter, M. Cardona, M. Dressel, B. Gorshunov, 
D. van der Marel, I. K{\"a}lin and A. Pimenov for useful discussions. 
This work was supported by the Los Alamos National Laboratory under the 
auspices of the U.S. Department of Energy and by the Swiss National 
Science Foundation through the National Center of Competence
in Research "Materials with Novel Electronic Properties-MaNEP".

\begin{appendix}

\section{Eigenmodes for alternating junctions } 

\label{appendixeigen}

We introduce the unit cell, which contains two different junctions 
and describe 
the system by the parameters $c_{ml},d_{ml},P_{ml}$ 
(cf. Eqs.~\ref{def1}), 
where $m$ denotes the unit cell and $l=1,2$ labels the junctions in the unit
cell. The equations inside the crystal are
analogous to Eqs. (\ref{e1})-(\ref{e3}), where the quasiparticle dissipation 
is taken into account by ${\tilde w}_l={\tilde \omega}_l^2/\omega^2_{c0,l}$
and ${\tilde \omega}_l^2 / \omega^2 = 1 - 4 \pi i \sigma_l \omega / 
\epsilon_{c0} \omega_{c0,l}^2 $:
\begin{eqnarray}
&&c_{m1}\eta^{-1}-d_{m1}\eta -c_{m-1,2}\eta +d_{m-1,2}\eta^{-1}=0, \nonumber \\
&&c_{m2}\eta^{-1}-d_{m2}\eta -c_{m,1}\eta +d_{m1}\eta^{-1}=0,  \label{1} \\
&&2(c_{m1}\eta^{-1}-c_{m-1,2}\eta)+(a-1)(P_{m1}-P_{m-1,2})+ \nonumber \\
&&i(\beta/b)(c_{m1}\eta^{-1}-d_{m1}\eta) =0 , \nonumber \\
&&2(c_{m2}\eta^{-1}-c_{m1}\eta)+(a-1)(P_{m2}-P_{m1})+ \nonumber \\
&&i(\beta/b)(c_{m2}\eta^{-1}-d_{m2}\eta) =0  , \\
&&P_{m1}({\tilde w}_1 -a)+\alpha(P_{m2}+P_{m-1,2}-2P_{m1})=  \nonumber \\
&&(c_{m1}+d_{m1})(1-2\alpha\beta)(\sin b/b),  \nonumber \\
&&P_{m2}({\tilde w}_2 -a)+\alpha(P_{m+1,1}+P_{m1}-2P_{m2})= \nonumber  \\
&&(c_{m2}+d_{m2})(1-2\alpha\beta)(\sin b/b). 
\label{2}
\end{eqnarray}

Using the Fourier transformation with respect to the discrete index $m$ we 
obtain the dispersion relation in the limit $b \ll q,\beta^{1/2}$, which is 
appropriate for oblique incidence in Sect. \ref{sectionalternate},
\begin{eqnarray}
&&(\nu^2+2 
\beta-\beta^2){\cal D}+4\alpha(a-1)(1-\nu^2)\beta +(a-1)^2\nu^2+ \nonumber \\
&&(\nu^2+\beta)(a-1)({\tilde w}_1 + {\tilde w}_2 
-2a-4\alpha)=0, \label{dis1a} \\
&&{\cal D}=({\tilde w}_1-a-2\alpha)({\tilde w}_2-a-2\alpha)
-4\alpha^2(1-\nu^2). 
\nonumber
\end{eqnarray}

In the opposite limit $q \ll b$ used for parallel incidence in
 Sect. \ref{sectionparallel}, we get the dispersion 
\begin{eqnarray}
&&{\rm Det} \left[ \hat{W}  \label{dispapp2}
\hat{\epsilon}-\frac{a-1}{2 + \beta} \hat{\Lambda} \right] =0, \\
&&{\hat \epsilon}_{ll}= \left[1-\omega_{0,l}^2 (1+2\alpha)/
\tilde{\omega}_l^2\right], \ \ {\hat \epsilon}_{12}={\hat \epsilon}_{21}
=2\alpha, 
\nonumber
\end{eqnarray}
where $\hat{W}$, $\hat{\epsilon}$ and $\hat{\Lambda}$ are matrices, 
$W_{ll} = {\tilde w}_l $,
$W_{12}=W_{21}=0$, and 
$\Lambda_{ll}= 1 + \beta$, $\Lambda_{12}=\Lambda_{21}=1$.

Using Eqs. (\ref{dis1a})  for $b \ll q,\beta^{1/2}$ 
and taking into account the lattice constant $2s$ in 
$q_p = 2s k_{z,p} $,  the refraction indices 
$n_p = c k_{z,p} / \omega = c \nu_p / s \omega $ of the bulk eigenmodes
for $\sigma_l=0$ are determined by ($z_0=a-1-2\alpha$)
\begin{eqnarray}
&&\nu_{1,2}^2(\omega)=(-P \pm \sqrt{P^2-16\alpha^2\beta Q})/8\alpha^2, \\
&&P(w)=w^2\delta-w(1+\delta)(1+2\alpha)+1+4\alpha (1-\beta z_0) , \nonumber \\
&&Q(w)=2w^2\delta-w(1+\delta)(1+a+4\alpha)+2a(1+4\alpha). \nonumber
\end{eqnarray}
Thereby and in the following the effect of dissipation can be included by
 replacing $w \rightarrow
{\tilde w}_1$  and $w \delta \rightarrow {\tilde w}_2$ and we will restrict
 the discussion to the case $\alpha^2>(a-1)\beta(1-\delta)/4$ and 
$\pi\sigma_c/\omega_{c0,1} \epsilon_{c0} < \alpha [(a-1)\beta]^{1/2} /
[ (1-\delta) (1+2\alpha) ]^{1/2}$, when an extremal point $w_e$ exists in the 
lower band, provided that $1 - \delta $ is of order unity. 
 
At $a=1$ we get $Q(w)=2P(w) + O (\alpha^2 \beta)$ and
$\nu_{1,2}^2$ is small near the zeros $w_{\rm low,up}$ of $P(w)$, which are 
given by Eq.~(\ref{100}). 
The reflection coefficient is determined predominantly by small 
$\nu_{1,2}^2(w)$, as in the case of identical junctions.  Therefore, in the 
following we will analyze the behavior of $\nu_{1,2}^2(w)$ by expanding 
around $w_{\rm low}$  ($w_{\rm up}$) for the lower (upper) band. 
With  $u_{\rm low,up}=w-w_{\rm low,up}$ 
we obtain $P(w)= \mp \lambda u_{\rm low,up}$ and 
$Q(w)=-[\pm 2\lambda + (a-1)(1+\delta)]u_{\rm low,up} \pm (a-1)L_{\rm low,up}$, 
where  we denote 
$\lambda=[(1-\delta)^2(1+4\alpha)+4\alpha^2(1+\delta)^2]^{1/2}$ and 
$L_{\rm low,up}=\pm[2+8\alpha-w_{\rm low,up}(1+\delta)]>0$
(upper/lower sign for lower/upper band).
From this the band edges $w_{\rm low,up}^{(\pm)}$
and special frequencies $w_{e,i}$ of the bands can be obtained in 
the limit $\sigma_l=0$, cf. Fig. \ref{schematicmixing2layer} and 
\ref{schematicnu2layer}. 

In the lower band positive real solutions for $\nu_{1}^2$ exist for 
$\sigma_l=0$ at $w_e=w_{\rm low}+u_{\rm low}
 < w < w_{\rm low}^{(+)} = 1 + 2 \alpha$, 
where $u_{\rm low}=[16\alpha^2\beta(a-1)L_{\rm low}]^{1/2}/\lambda$.  
At the extremal 
point $w_e$ we get $\nu_1^2=\nu_2^2=[\beta(a-1)L_{\rm low}]^{1/2}/2\alpha$, 
while 
the upper edge $w_{\rm low}^{(+)}$ is determined by the condition 
$\nu_1^2=1$ by noting that $P(1+2\alpha)=-4\alpha^2$. The value 
$\nu_2^2$ is positive at $w\leq w_0 = w_{\rm low} + (a-1)L_{\rm low}
 /(2\lambda + (a-1) (1+ \delta) ) $ and approaches $-2\beta$ for
 $w>w_0$ till the second 
band is reached. In the following we consider the
case $w_{\rm low}^{(+)} > w_0$ and hence the upper 
edge of the lower band is $w_{\rm low}^{(+)}$. 
In the range $[w_e, w_0]$ two propagating 
modes with normal and anomalous dispersion exist, while for
$w \in [w_0, w_{\rm low}^{(+)}]$, 
$\nu_1$ is propagating and $\nu_2$ decaying, which 
is very similar to  a system with identical layers. Also note that the width 
$w_0 - w_e$ of the resonance in $T(w)$ is not proportional to the $c$-axis
 coupling $\alpha$, but is mainly given by the angle of incidence. 

In contrast to this, the behaviour of $\nu_{1}^2$ and  $\nu_{2}^2$ in the
upper band is quite different because the dispersion here is anomalous at 
any frequency. 
In this range $\nu_2^2$ is negative and $|\nu_2^2|>2\beta$ and the band edges
are determined by the conditions $\nu_{1}^2=0$ or $\nu_{1}^2=1$ respectively.
The value $\nu_1^2$ is positive inside the band $w_{\rm up}^{(-)} 
< w< w_{\rm up}^{(+)}$, where 
$w_{\rm up}^{(-)}=(1+2\alpha)/\delta$ and 
$w_{\rm up}^{(+)}=w_{\rm up} + u_{\rm up} $ where 
$ u_{\rm up} = - (a-1)L_{\rm up} /(-2\lambda + (a-1) (1+ \delta)) >0$. 
At the point $w=w_i=w_{\rm up}$ we obtain $-\nu_2^2=\nu_1^2=
[\beta(a-1)L_{\rm up}]^{1/2}/2\alpha$, which corresponds to the 
frequency of maximal transmission according to Eq.~(\ref{tmaxi}).

Similarly to Eqs. (\ref{eqc})-(\ref{eqA}) we make an ansatz for the 
bulk eigenvectors 
\begin{eqnarray}
c_{ml}&=& \sum_{p=1,2}\gamma_{p} c_l(q_p)  \exp(iq_pm), \label{11}\\
d_{ml}&=&\sum_{p=1,2}\gamma_{p}d_l(q_p)\exp(iq_pm), \\
P_{ml}&=&\sum_{p=1,2}\gamma_{p}P_l(q_p)\exp(iq_pm) . 
\label{50}
\end{eqnarray}
Using Eqs.~(\ref{1})-(\ref{2}) we obtain the coefficients 
($\underline{P}=(P_1(q), P_2(q))$) 
\begin{eqnarray}
c_1&=& 1, \ \  c_2(q)  =    
  \frac{\alpha(a-1)(1+e^{iq})}{{\cal D}+(a-1)({\tilde w}_2-a-2\alpha)} ,
\label{coeffc2} \\
d_1 &=&\frac{1-[1+ (\eta^2 - \eta^{-2}) c_2(q)]
e^{-iq}}{\eta^2-\eta^{-2} e^{-iq}}, \label{coeffd1} \\
d_2 &=& 
\frac{\eta^{-2} - \eta^2 + (1-e^{-iq}) c_2(q)}{\eta^2-\eta^{-2}e^{-iq}}, 
 \label{coeffd2} \\ 
\underline{P} &=& \underline{\underline{M}} 
\left( \underline{c} + \underline{d} \right) ,  \label{Avect} \\ 
\underline{\underline{M}} &=& \frac{1}{\cal D}
\left( 
\begin{array}{cc}
 { \tilde w }_2 - a - 2 \alpha  &  - \alpha ( 1 + e^{-iq} )   \\
 - \alpha ( 1 + e^{iq} ) & {\tilde w}_1 - a - 2 \alpha  
\end{array} 
\right) . 
\end{eqnarray}
Here $c_2(q)$ is given in leading order in $b$ and $\beta^{1/2}$. 
In order $O(b^0)$ we get $d_1=c_1=1$ and $d_2=c_2(q)$. 

To determine $\gamma_{1}/\gamma_{2}$ we use 
the microscopic boundary condition for the surface junction (analogous to 
Eq.~(\ref{mbc})). Near $w_{e,i}$ we get in leading order in $\beta^{1/2} 
\sim \epsilon_a/n^2_p$
\begin{eqnarray}
&&P_{01}({\tilde w}_1 -a)+\alpha\{P_{02}
-[1+a+(a-1)\epsilon_{c0}]P_{01}\}= \nonumber \\
&&(c_{01}+d_{01})[1+\alpha(\epsilon_{c0}+1)]. \label{mbc2}
\end{eqnarray}
Again we simplify this equation by subtracting the (hypothetical) bulk
 equation for $P_{01}$, 
which follows from Eq.~(\ref{2}) with $P_{m=-1,2}$ given by Eq.~(\ref{50}), 
and the real surface Eq.~(\ref{mbc2}) for $P_{01}$: 
\begin{equation}
P_{m=-1,2} = \sum_p \gamma_{p} P_2(q_p) \exp(-iq_p)=0 . 
\label{111a}
\end{equation}
Note that this ABC  has only been derived in leading order in 
$\epsilon_a/n^2_p$ and near the resonance frequencies $w_{e,i}$. 

\section{Reflectivity in parallel incidence}

\label{appendixparallel}

For arbitrary $\beta_0$ and $\alpha$ the solutions of the dispersion 
Eq.~(\ref{dispapp2}) in the case $q \ll b$ are given by:
\begin{eqnarray}
\frac{c^2 k_{xp}^2}{\omega^2 \epsilon_{c0}} &=& 
\frac{ (w \delta ( w - w_{\rm pole})  + i S_1 ) (1+\beta_0) 
}{\beta_0 [ w^2 \delta - 2 \alpha w (1 + \delta ) + i S_2 ] } 
\label{kxgen} 
\\ 
\nonumber &&
  \left[ 1 \pm 
\left( 1 - \frac{K}{ (w \delta ( w - w_{\rm pole})  + i S_1 )^2 (1+\beta_0)^2 
   } \right)^{1/2}
 \right] \\
K &=& 2 \beta_0 (w^2 \delta - 2 \alpha w (1+\delta)+ i S_2) \\
  &&   ( \delta (w-w_{\rm low}) (w - w_{\rm up}) + i S ) (1 + \beta_0 /2)  
\nonumber \\
S&=&w^{1/2} [(2\alpha+1) \delta w (\tilde{\sigma}_1+\tilde{\sigma}_2)-
(1+4\alpha)(\tilde{\sigma}_1+\tilde{\sigma}_2 \delta)],
\nonumber \\
S_1&=&w^{3/2} \delta (2 \alpha + 1/2) ({\tilde \sigma}_1 + 
{\tilde \sigma}_2), \nonumber \\
S_2 &=& 2 \alpha w^{3/2} \delta ({\tilde \sigma}_1 + {\tilde \sigma}_2) . 
\nonumber 
\end{eqnarray}
Away from the pole $w_{\rm pole}$ in ${\tilde \epsilon}_c$ we obtain in
leading order in $\beta_0$ for arbitrary $\alpha$ 
\begin{eqnarray}
&&1-\frac{1}{a_1}=\frac{c^2k_{x1}^2}{\omega^2\epsilon_{c0}}=
\frac{\delta (w-w_{\rm low})(w -w_{\rm up})+ i S}{w \delta (w-w_{\rm pole})+i
 S_1 - \beta_0 c_0}, \\
&& 1-\frac{1}{a_2}= \frac{c^2k_{x2}^2}{\omega^2\epsilon_{c0}}
=\frac{ (w \delta ( w - w_{\rm pole} ) + i S_1) (1+\beta_0)}{
(\beta_0/2) ( w^2 \delta -2 \alpha w (1+\delta) + i S_2)}. 
\end{eqnarray}

In the ansatz Eq.~(\ref{by}) for the field $B_y$ outside the crystal
 the continuity equation at $x=0$ gives 
for ${\cal B}_y(k_z)$ the following expression
\begin{equation}
{\cal B}_y(k_z)= \frac{\epsilon_{c0} \omega}{ck_x}
\int dzG(z)\exp(-ik_zz), 
\end{equation}
where for $2ms\leq z <(2m+1)s$ we obtain ($g_p = \omega \epsilon_{a0}^{1/2} /
[c a_p^{1/2}]$)
\begin{equation}
G(z)= \sum_{p=1,2}  c_1^{(p)} \exp(ig_p z)+d_1^{(p)}\exp(-i g_p z), \nonumber
\end{equation}
and for $(2m+1)s\leq z <(2m+2)s$ analogously with
$c_1^{(p)} \rightarrow c_2^{(p)}$ and $d_1^{(p)} \rightarrow d_2^{(p)}$.

We derive $k_z=\pm g_p+(\pi/s)j$, where $j$ is an integer. 
For nonzero $j$ we obtain $k_z\geq \pi/s$ and hence $k_{xp}$ is imaginary
 with large $|k_{xp}|$.
For $j=0$ we obtain $\langle {\cal B}(g_p) \rangle=\langle 
{\cal B}(-g_p) \rangle=0$ by averaging over the two junctions in the unit 
cell, as  $c_1^{(p)}+c_2^{(p)}=d_1^{(p)}+d_2^{(p)}=0$  with accuracy 
$b/\beta_0 \sim \lambda_{ab} / \lambda_c \ll 1$ from Eqs.~(\ref{1})-(\ref{2}).
As a result, in Eq.~(\ref{by}) the 
terms with amplitudes ${\cal B}_y (k_z)$ may be dropped. Then the amplitude 
of the reflected wave  $B_{y}^{\rm ref}$
is determined by Eq.~(\ref{kappagen}), where the averaged
 magnetic and electric fields at the boundary $x=0$ are 
\begin{eqnarray}
&&\langle B_{y}\rangle=-\frac{c}{2\omega}\sum_{p=1,2}a_pk_{xp}
[P_1^{(p)}+P_2^{(p)}], \\
&&\langle E_{z}\rangle=\frac{1}{2} \sum_{p=1,2}a_p[P_1^{(p)}+P_2^{(p)}].
\end{eqnarray}
These equations lead to the 
reflection coefficient, Eq.~(\ref{kappagen}), where 
\begin{equation}
\kappa_{\parallel}= \sqrt{\epsilon_{\parallel}^{\rm eff}} = 
\frac{a_1+a_2Z}{a_1n_{x1}+a_2n_{x2}Z}, \ \ 
Z=\frac{P_1^{(1)}+P_2^{(1)}}{P_1^{(2)}+P_2^{(2)}}
\end{equation}
with $n_{xp} = c k_{xp} / \omega$. As $a_2  \ll a_1$ for 
$w \neq w_{\rm pole}$ the conventional Fresnel expression is valid everywhere 
except at the upper edge $w_{\rm pole} $ of the lower band. 

To determine $Z$, we use the additional boundary condition $P_x=0$ of the 
Pekar type in the form 
\begin{eqnarray}
&&E_x(x=0)=\sum_{p=1,2} \frac{1}{k_{xp} a_p^{1/2}}
(c_1^{(p)}\eta_p^{-1}-d_1^{(p)}\eta_p)=0 . 
\nonumber 
\end{eqnarray}
From this and the relation $c_m^{(p)}=d_m^{(p)}$ the condition 
$\sum_p c_1^{(p)} / [k_{xp} a_p ] =0 $ follows. 
To express it in terms of $P_1^{(p)}+P_2^{(p)}$ we derive from 
Eqs.~(\ref{1})-(\ref{2}) 
\begin{eqnarray}
c_1^{(p)} &=&-(1/4)(a_p-1)P_1^{(p)}(1-f_p), \nonumber \\
f_p&=&\frac{P_2^{(p)}}{P_1^{(p)}} =    \nonumber
\frac{({\tilde w}_1-1-2\alpha) (2 a_p + \beta_0) -
(a_p-1)(a_p+\beta_0)}{(a_p-1)(a_p+\beta_0)-2 \alpha (2 a_p + \beta_0) }.
\nonumber
\end{eqnarray}
Finally we obtain 
\begin{equation}
Z=-\frac{(1-f_1)(1+f_2)(a_1-1)k_{x2} a_2 }
{(1-f_2)(1+f_1)(a_2-1)k_{x1} a_1}. 
\end{equation}

\end{appendix}


\end{document}